\documentclass[aps,prd,floatfix,preprintnumbers,altaffilletter,superscriptaddress,reprint,longbibliography,twocolumn,nofootinbib]{revtex4-2}
\usepackage{graphicx,url,amssymb,amsmath,rotating,color,units,wasysym,epsfig,multirow,epstopdf, stackengine}
\usepackage{cancel}
\usepackage{physics}
\usepackage{appendix}
\usepackage{amssymb}
\usepackage[colorlinks,urlcolor=blue,citecolor=blue,linkcolor=blue]{hyperref}
\usepackage[normalem]{ulem}
\usepackage{tabularx}
\newcolumntype{Y}{>{\centering\arraybackslash}X}
\usepackage{braket}
\usepackage{cleveref}
\usepackage{ushort}

\usepackage{soul}

\usepackage[commentmarkup=uwave]{changes}
\definechangesauthor[name=Ethan, color=cyan]{EP}

\newcommand{\fE}{\ensuremath{e}}
\newcommand{\pN}{\ensuremath{N}}
\newcommand{\qv}[1]{\ensuremath{\hat{#1}}}
\newcommand{\rv}[1]{\ushort{#1}}
\newcommand{\qrv}[1]{\ensuremath{\ushort{\hat{#1}}}}

\newcommand{\CIT}{\affiliation{TAPIR, California Institute of Technology, Pasadena, CA 91125, USA}}
\newcommand{\CITLab}{\affiliation{LIGO Laboratory, California Institute of Technology, Pasadena, California 91125, USA}}
\newcommand{\IQIM}{\affiliation{Institute of Quantum Information and Matter, California Institute of Technology, Pasadena, California 91125, USA}}

\begin{document}

\title{{Photon temporal-mode readout for inference of neutron star merger remnant gravitational waves}}

\author{Ethan Payne}
\email{epayne@caltech.edu}
\CIT
\CITLab

\author{Lee McCuller}
\email{mcculler@caltech.edu}
\CIT
\IQIM
\CITLab

\author{Katerina Chatziioannou}
\email{kchatziioannou@caltech.edu}
\CIT
\CITLab

\begin{abstract}
Gravitational waves emitted after neutron star binary coalescences and the information they carry about dense matter are a high-priority target for next-generation detectors.
Even though such detectors are expected to observe millions of signals, detectable postmerger emission will remain rare.  
In this work, we explore postmerger detectability and inference through an alternative detector readout scheme for data dominated by quantum-noise, which is the case above $1$\,kHz: photon-counting. 
In such a readout, signals and noise become quantized into discrete distributions corresponding to the detection of single photons measured in a chosen basis of modes. 
Through simulated data, we demonstrate that photon counting can be efficient even for weak signals.
We find ${\sim}1$ in 100 signals with a postmerger signal-to-noise ratio of 0.2 can result in a single photon and thus be detected. 
Furthermore, after $2\times10^4$ signals --- equivalent to $10^{-2}$ to $1.5$ years of observation --- photon counting results in a twofold improvement in the measurement of the radius of a $1.6\,M_\odot$ neutron star.
Constraints can be further tightened if the detector classical noise is reduced.
Photon counting offers a promising alternative to traditional homodyne readout techniques for extracting information from low signal-to-noise ratio postmerger signals.
\end{abstract}

\maketitle

\section{Introduction}

Future generations of gravitational-wave (GW) detectors will enhance studies of dense nuclear matter~\cite{Chatziioannou:2024tjq} by providing access to the postmerger signal emitted from binary neutron-star (BNS) mergers~\cite{Clark:2015zxa, Evans:2021gyd, Breschi:2022ens}.
Direct observations of the emission from the merger remnant will provide information about the dense-matter equation-of-state (EoS)~\cite{Chatziioannou:2017ixj,Easter:2018pqy, Tsang:2019esi,Chatziioannou:2020pqz,Wijngaarden:2022sah,Breschi:2024qlc,Huez:2025gja} (including the possibility of phase transitions~\cite{Most:2018eaw,Bauswein:2018bma}), the nature of hyper and supramassive NSs~\cite{Beniamini:2021tpy}, and the production of electromagnetic radiation~\cite{Li:1998bw, Metzger:2010}. 
The density and temperature regime accessible to such observations cannot be reached by other probes~\cite{Miravet-Tenes:2024vba,Gieg:2025ivb,Raithel:2025geu,Villa-Ortega:2023cps,Breschi:2021xrx}.

postmerger signals are emitted at frequencies from $1.5$\,kHz to $4$\,kHz, and redshifted down to $500$\,Hz to $4$\,kHz~\cite{Haster:2020sdh} depending on the binary distance.
For ground-based detectors such as LIGO~\cite{LIGO}, Virgo~\cite{Virgo}, KAGRA~\cite{KAGRA} and future observatories such as the Cosmic Explorer (CE)~\cite{Evans:2021gyd} and Einstein Telescope~\cite{Branchesi:2023mws}, sensitivity at these frequencies is limited by quantum measurement noise from photon shot noise on the interferometer recorded fringe light~\cite{Caves:1981}. 
The quantum shot noise is now being reduced by almost 6\,dB by injecting squeezed light~\cite{Lough:2020xft, Ganapathy:2023, Capote:2024rmo}, and future observatories plan to achieve 10\,dB of quantum noise suppression~\cite{Evans:2021gyd, Branchesi:2023mws}, but will be limited to around 8\,dB above 1kHz even with optimistic assumptions on internal losses\cite{MiaoPRX19QuantumLimit}.

Despite these improvements, most postmerger signals are still expected to be weak~\cite{Evans:2021gyd, Torres-Rivas:2018svp}.
Next-generation detectors are only expected to record postmerger signals with a signal-to-noise ratio (SNR) greater than five approximately once per year~\cite{Evans:2021gyd}. 
Below this SNR threshold, any signal will also yield almost entirely uninformative posteriors on the source properties~\cite{Chatziioannou:2017ixj}. 
Given that CE is projected to observe hundreds of thousands of BNS mergers annually~\cite{Baibhav:2019gxm, Iacovelli:2022bbs, Gupta:2023lga}, techniques that lower the threshold for postmerger detection and informativeness will substantially increase the scientific yield.

Future observatories are assumed to employ the same measurement process as existing detectors: recording modulations of fringe light and calibrating these modulations to a strain time-series~\cite{LIGOScientific:2010weo,Sun:2020wke}. 
In the quantum measurement literature, this measurement process is known as \emph{homodyne detection}. {In the GW instrumentation community, homodyne detection is employed when using fringe light as a local oscillator (also called DC readout) or when using a separately-supplied local oscillator for balanced homodyne detection~\cite{fritschel2014balanced}.}
Detector sensitivity with any form of homodyne readout is quantified through the power spectral density (PSD) of the strain channel, where the total density is the sum of quantum and classical noise contributions.
Both the time-series data product and the quantum shot noise are a result of using homodyne readout, {regardless of whether it is implemented through DC readout or balanced homodyne detection}.

The quantum noise is unlike classical noises, because its effect depends on the choice of quantum measurement process. 
Alternative measurements schemes are physically possible and modify the statistical properties of both signals and noise. 
Such readout methods are still limited by randomness from quantum mechanical indeterminacy, but can have more favorable statistical figures of merit than those implied by additive quantum shot noise power, even with squeezing applied~\cite{Ng:2016, Tsang:2023, Gardner:2024frf}.
Additionally, they can be restricted to the frequency bands dominated by quantum noise, in practice greater than $1\,$kHz, resorting to standard homodyne elsewhere.

In this exploratory study, we assume the eventual existence of hardware that performs matched-filtering directly on the electromagnetic field that is output by the interferometer.  
We choose a quantum measurement of power rather than of amplitude which is \emph{discretized into photon counts}~\cite{McCuller:2022hum}. 
We present the statistical framework and observational methodology, applied to BNS postmergers.
We also compare to the standard time-series readout that provides a continuous measure of electromagnetic amplitude through homodyne.

{A heuristic argument of the potential benefits photon counting may provide follows. 
If postmerger inference requires marginalization over nuisance parameters (such as the waveform phase that is unique for each BNS event), then the statistic to ``stack'' many events has a poor scaling with the event number, due to the similarity with being an an excess power search~\cite{YangPRD18GravitationalWave}. This poor scaling can be justified using Fisher Information, see App.~\ref{app:FI}. Under photon counting, the rate of photons produced by the signal scales inversely to the power spectrum of shot noise, due to common calibration factors between the photon emission rate and noise process~\cite{McCuller:2022hum, Vermeulen:2024vgl}. 
In the regime where classical noise is small, then the only mechanism to produce a photon at each event is from the signal, and thus recording even a single photon {can result} in a postmerger detection. By identifying the exact frequency of the photon, the postmerger frequency can also be constrained, as classical noise is too low to produce a background that confounds parameter estimation and shot noise only determines the rarity of seeing a signal photon, but does not create fluctuations that could cause false-positive detections. The detection rate under photon counting thus {could have a fundamentally better scaling} than an excess power search under homodyne. In reality, the constants of proportionality are important, the classical noise is nonzero, and the complete inference processes for each readout must be compared. The potential for improved detection and the complexity of the above statistical argument motivates this numerical study.}

We implement a full {Bayesian} analysis of simulated signals, directly comparing photon counting and homodyne.
In Sec.~\ref{sec:pc}, we outline the photon counting methodology and translate it into the standard formulation  of GW inference. 
In Sec.~\ref{sec:single}, we consider individual signals and address some of the quirks of inference which relies on discrete observational data, as opposed to the continuous output of homodyne readout. 
We find that photon counting measurements are able to occasionally gain meaningful information from observations with SNR $ \lesssim 1$. 
Then, in Sec.~\ref{sec:hier} we consider EoS constraints from a population of sources. 
In a hierarchical context, photon counting performs comparably to homodyne readout with 10\,dB of squeezing, under current projections about the level of classical noise. 
An order-of-magnitude reduction of classical noise allows for significantly improved constraints. 
Implications of the construction of such a readout scheme and concluding remarks are presented in Sec.~\ref{sec:imp}. 
For the remainder of the manuscript, and while this analysis can be extended to multiple detectors, we consider a single CE interferometer {using fused silica mirrors, 1064\,nm technology, and 1.5\,MW of arm power. CE in this configuration is projected to have 10\,db of squeezing at most frequencies, but at high frequencies only achieves 8\,db due to internal losses~\cite{MiaoPRX19QuantumLimit}.}

\section{Photon Counting for postmerger detection and inference}
\label{sec:pc}

In this section, we present photon counting as an alternative readout scheme. 
The main output of this section is the likelihood function for photon counting, Eq.~\eqref{eq:likelihood}, that replaces the standard Gaussian likelihood of homodyne in inference.
We begin with a summary of the more familiar homodyne readout in Sec.~\ref{subs:homodyne}. 
We then detail the photon counting readout and its statistics in Secs.~\ref{subs:rb} and~\ref{subs:pc}.  
Finally, in Sec.~\ref{subs:pc_like}, we outline the full likelihood function describing photon counting.

Throughout, a `hat' indicates a quantum operator, as opposed to complex numbers. Statistical random variables that are not operators are indicated by an underline.\footnote{{The underline is to remind of the more typical notation of using capital lettering for random variables. In our notation, capitals signify physical units of energy, power, or number rather than field or signal amplitude.}}

Quantum operators represent measurable quantities, but their different parts are not individually measurable.
Statistical random variables can be treated similarly to quantum operators with the property that they commute with all other operators and the expectation value of their first moment is zero.
Standard addition functions apply between operators, random variables, and complex numbers.

\subsection{Summary of the homodyne readout}\label{subs:homodyne}

We begin with a description of interferometric GW detection and the statistical framework for homodyne readout.
Michelson interferometers are generally described as directing laser light to a beamsplitter, where it splits and travels down two arms to accumulate a phase shift, reflects at respective end-mirrors, and then is recombined upon a second pass through the beamsplitter. 
The light fields are combined into constructive and destructive interference outputs. 
Advanced LIGO operates a DC homodyne readout~\cite{fritschel2014balanced} where, even in the absence of differential length, a small static offset from destructive interference is maintained at the output port.
This static fringe offset both enables signal readout and determines the form of quantum noise.

The differential phase accumulation is encoded in the optical field $\rv{\fE}_h(t)$ which carries the strain signal, $h(t)$, and classical noise sources, $\rv{n}(t)$.
Classical noise are essentially all noise sources that are not inherent to the nature of the readout scheme itself: seismic, thermal, electronics, etc.~\cite{Cahillane:2022pqm, Capote:2024rmo}. 
Both $h(t)$ and $\rv{n}(t)$ are calibrated from strain into optical field by applying an optical detector gain, $g(f)$, expressed in the frequency-domain as a linear time-invariant filter. 
GW detectors use recycling cavities optimized for a bandwidth of $\Delta F_{\det}$ (typically around 450\,Hz~\cite{LIGO}) and circulating arm power $P_{\text{arm}}$ (around 400\,kW in O4 LIGO~\cite{Capote:2024rmo} to 1.5\,MW in future detectors~\cite{Evans:2021gyd, Branchesi:2023mws}). 
An idealized, but accurate, expression for the optical calibration function is~\cite{Rakhmanov:2008is}
\begin{equation} \label{eq:cal_gain}
    g(f) = i k \sqrt{\eta \frac{c L P_{\text{arm}}}{\pi \Delta F_{\text{det}}}}\left(1 + \frac{if}{\Delta F_{\text{det}}}\right)^{-1}\,,
\end{equation}
where $k$ is the wave number of the interferometer's source laser, $2\pi/1064$\,nm for LIGO and CE~\cite{LIGO, Evans:2021gyd}, $c$ is the speed of light, $L$ is the arm length, and {$\eta$ is the readout optical efficiency in power units. We take $\eta = 1$ for simplicity but include it to show the simple scaling and impact of efficiency on counting.}
The phase-factor of $i$ applies the convention that signals are imprinted in the light-field phase.
Applying the calibration gives the frequency-domain expression for the optical field {component sourced by signal and classical noise.}
\begin{equation}
    \rv{\fE}_h(f) = g(f) \left[h(f) + \rv{n}(f)\right]\,.~\label{eq:E_h}
\end{equation}
{The symbol $e$ above and in following expressions represents a one-dimensional traveling-wave field amplitude that is proportional to the electric field; it is in units of field flux, $\sqrt{\text{photons}/\text{s}}$.}

The classical noise, $\rv{n}(f)$, is a zero-mean random variable with a PSD given by
\begin{equation}
  \braket{\rv{n}(f) \rv{n}^*(f')} = \frac{1}{2}S_n(f)\delta(f-f')\,,
\end{equation}
where the angled brackets indicate the expectation over noise realizations {and $\delta(f-f')$ is the Dirac delta function}.

The total output field at the destructive interference port is a sum of the time-varying differential-arm component, the static fringe offset field, and the quantum operator for the field reflecting from the signal port of the interferometer\footnote{The given formulas and interpretation of the quantum operator are used to calculate shot noise. Radiation pressure, the other type of quantum noise, can be treated similarly~\cite{Caves:1981}, but are omitted in this treatment.}~\cite{Caves:1981},
\begin{equation} \label{eq:e_out_HD}
  \qrv{\fE}_{\text{out}}(t) = \rv{\fE}_{\text{h}}(t) + \fE_{\text{fringe}} + \qv{\fE}_q(t)\,.
\end{equation} 
The static fringe field, $\fE_{\text{fringe}}$, is a small constant offset from destructive interference. It enables measurement of $\rv{\fE}_{\text{h}}(t)$ via small modulations in the power.
The quantum term $\qv{\fE}_q(t)$ is an operator, and its inclusion promotes $\qrv{\fE}_{\text{out}}(t)$ into an operator as well. Physically, $\qrv{\fE}_{\text{out}}$ fully represents the quantum state of the light that is coming out of the interferometer.\footnote{These formulas are using a Heisenberg formulation, so the quantum state is encoded into the operators.}
The term $\qv{\fE}_q(t)$  represents the quantum vacuum state (zero-point fluctuations in the field with zero mean but nonzero variance) that is injected into the interferometer through the destructive interference (dark) port. The state represented by the Heisenberg-picture operator $\qv{\fE}_q(t)$ might be squeezed.

{The interferometer also transforms the input states, e.g., through radiation pressure interactions. The simple linear inclusion of the input quantum operator into the output operator of Eq.~\eqref{eq:e_out_HD} indicates that we are not considering further transformations by the interferometer in this work~\cite{Caves:1981}, as they only affect quantum states at frequencies too low to be relevant for BNS postmerger signals.}

The operator for power at the output port\footnote{The operator ordering is understood through the Glauber theory of photodetection~\cite{PhysRev.130.2529}.} is then
\begin{equation}
    \qrv{P}_{\text{out}}(t) = \qrv{\fE}^\dagger_{\text{out}}(t)\qrv{\fE}_{\text{out}}(t) = P_{\text{fringe}}  + \qrv{P}_{\delta}(t)+\qrv{P}_\epsilon(t) \,,
\end{equation}
which is factorized into
\begin{align}
  P_{\text{fringe}} &= |{\fE}_{\text{fringe}}|^2\,, \\
    \qrv{P}_{\delta}(t) &=  2\Re\{\rv{\fE}_{\text{h}}(t)\cdot \fE^*_{\text{fringe}}\} + \qv{P}_q(t)\,,~\label{eq:deltaP}
                         \\
\qv{P}_q(t) &= \qv{\fE}_{q}(t)\cdot \fE^*_{\text{fringe}} + \qv{\fE}^\dagger_{q}(t)\cdot \fE_{\text{fringe}}\,, \\
  \qrv{P}_\epsilon(t) &= |\rv{\fE}_{h}(t)|^2 + \qv{\fE}^\dagger_q(t)\qv{\fE}_q(t) +\qv{\fE}^\dagger_q(t)\rv{\fE}_h(t) {+}  \rv{\fE}^*_h(t)\qv{\fE}_q(t)\label{eq:Pepsilon}\,,
\end{align}
The term $P_{\text{fringe}}$ is the constant static fringe power.
The term $\qrv{P}_{\delta}(t)$ is proportional to the fringe field and would thus be absent under perfect destructive interference.
Its first component encodes the modulation of the fringe field due to the differential-arm field and enables measurement of the latter as it is the only non-negligible term that contains $\rv{\fE}_{\text{h}}(t)$.
The second component is the zero-mean quantum operator for the quantum noise process, and is the mathematical representation of the fact that quantum noise is inherent to the type of readout scheme.
The term $\qrv{P}_\epsilon(t)$ is minuscule compared to the other terms and can be ignored for homodyne readout.

Expressing the signal in terms of strain and converting to the frequency domain yields
\begin{equation}
  \qrv{P}_\delta(f) = C(f) \left[h(f) + \rv{n}(f) + e^{-z} \qv{q}(f)\right]\,, \label{eq:HD_terms}
\end{equation}
where ${C(f) = 2g(f)\sqrt{P_{\text{fringe}}}}$.
The term $e^{-z} \qv{q}(f)$ represents the quantum noise-producing operator in units of strain. {Here $\qv{q}(f)$ is an operator for the vacuum} that results in shot-noise and the factor of $e^{-z}$ captures improvement for squeezing. 
The expectation value for the spectral density is~\cite{Caves:1981}
\begin{equation}
  S_{P_q}(f) = e^{-2z} 2\hbar k c P_{\text{fringe}}\,.
\end{equation}
Physically, this noise corresponds to the shot noise from Poisson statistics of measuring the fringe offset photons. 
Improved (reduced) noise from injecting squeezed states is expressed through the parameter $e^{-2z}=10^{-{\text{dB}_{\text{sqz}}}/10}$, where ${\text{dB}_{\text{sqz}}}$ is in decibels of observed quantum noise-power reduction.
The shot noise PSD without squeezing and calibrated in units of strain is~\cite{Ganapathy:2023},
\begin{equation}
  S_{q}(f) = \frac{S_{P_q}(f)}{\abs{C(f)}^2} = \frac{\hbar k c}{2 |g(f)|^2}\,.
  \label{eq:shotnoisePSD}
\end{equation}
The principal difference between $S_{q}(f)$ and $S_{P_q}(f)$ is the choice of units and the (non) inclusion of squeezing. The latter is in physical or low-level interferometer measured-power units, while the former is calibrated to strain spectral density. 
After calibration, shot noise no longer depends on $P_{\text{fringe}}$, and only depends on the amount of power in the interferometer arms, the interferometer bandwidth, and the squeezing level. 
We choose to omit the squeezing level in $S_q(f)$ for later convenience, following Eq.~\eqref{eq:inner_pc}, {in using the quantum spectral density rather than the optical gain as it is more typically available.}
Although shot noise is Poissonian in nature, the large number of photons in $P_{\text{fringe}}$ allows for it to be approximated as a Gaussian noise process, with an independent normal distribution at every frequency.

Altogether, the strain measured using the fringe light as a homodyne readout follows from Eq.~\eqref{eq:HD_terms},
\begin{equation}
    \qrv{h}_{\textsc{hd}}(f) = h(f) + \rv{n}(f) + e^{-z} \qv{q}(f)\,,
\end{equation}
with a noise background {factorized into classical and quantum terms} by
\begin{equation}
  S_{\textsc{hd}}(f) = S_n(f) + S_q(f) \cdot 10^{-{\text{dB}_{\text{sqz}}}/10}\,.
\end{equation}
The astrophysical strain, $h(f)$, and two separate noise processes sum to the measured strain $\rv{h}_{\textsc{hd}}(f)$. 
The first noise process is classical noise with strain $\rv{n}(f)$ and spectrum $S_n(f)$ which arises from random thermal, mechanical, electrical and laser processes in the interferometer~\cite{Cahillane:2022pqm}. 
The second noise process is quantum noise, which results from the Poissonian arrival of photons from the fringe light~\cite{Caves:1981}.
Combining the strain and PSD results in the standard Gaussian likelihood for the observed data given a model for the GW signal parametrized by $\theta$, $h_{\textrm{sig}, \theta}(f)$~\cite{Romano:2016dpx} as well as a corresponding noise-weighted inner product and optimal SNR definition of
\begin{align}\label{eq:snr_olap}
  (x | y)_{\textsc{hd}} &\equiv 2\int_{-\infty}^{\infty} \dd f \frac{x(f) y^*(f)}{S_{\textsc{hd}}(f)}\,,\\
  \textsc{Snr}[h_{\mathrm{sig},\theta}(f)] &\equiv \sqrt{\big(h_{\mathrm{sig},\theta} | h_{\mathrm{sig},\theta}\big)_{\textsc{hd}}}\,.\label{eq:snr}
\end{align}

\subsection{Readout Basis}~\label{subs:rb}

Before describing photon counting, we first introduce the concept of a temporal-mode basis, with which both homodyne and photon counting can be described and contrasted.
An astrophysical event is described by a waveform $h_{\mathrm{sig}, \theta}$ with parameters $\theta$.
Inference amounts to the posterior distribution for $\theta$ given measured data using some measurement scheme.
Signals can be decomposed into a vector of individually measurable and physically independent coefficients for each element of some basis.
We describe a basis of temporal modes with a set of $M$ orthogonal templates, $d_k(f)$, that are normalized to unit strain energy
\begin{align}
  \delta_{kj} = \int_{-\infty}^{\infty} \dd f d^*_k(f) d_j(f) \equiv \big(d_k|d_j\big)_{\textsc{tm}}\,,
\end{align}
where $\delta_{kj}$ is the Kronecker delta and we define an unweighted inner product.  
Each template has a central frequency
\begin{equation}
f_k = \int_{0}^\infty \mathrm{d}f\, f |d_k(f)|^2\,,\label{eq:template-fk}
\end{equation}
and bandwidth
\begin{align}
  \Delta f_k &= \sqrt{\int_{0}^\infty f^2 |d_k(f)|^2\mathrm{d}f\, - f_k^2}\label{eq:template-dfk}\\
  &\approx \max_f |d_k(f)|^{-2} \approx |d_k(f_k)|^{-2}\,.
\end{align} 
The basis templates can be chosen arbitrarily and should be optimized for the application. 
In practice, the basis is finite, so linear combinations of its elements may only approximate the full waveform. 
For waveforms with a finite support in time $\Delta T$ and frequency $\Delta F$, the Nyquist theorem establishes that at most $2\Delta T\Delta F$ independent basis components can be orthonormal over that area, setting a bound on the necessary $M$.

With such a basis definition, we can reframe homodyne. 
In a homodyne readout, data are measured in the form of a sampled timeseries of real values that is classically convertible into a time, frequency, or arbitrary basis.
The homodyne measures ${P_j = \qrv{P}_{\mathrm{out}}(t_j)}$ to produce $\qrv{h}_{\textsc{hd}}(f_j)$, discretized in time and frequency. 
The use of sampling in an analog-to-digital converter is a choice of basis that is orthogonal by having nonoverlapping support in the time domain. 
Given this choice, we can classically process the homodyne readout to produce the set of coefficients $\int_{-\infty}^{\infty}\dd f d^*_k(f) \qrv{h}_{\textsc{hd}}(f)$.

Photon counting relies on a different set of temporal-mode basis with critical differences.
First, the measurement is not a continuous field amplitude, but rather a discrete set of photons counted in each mode.
Second, the basis is physically implemented in hardware, rather than being a classical postprocessing choice.
This means that, third, the measurement occurs in each basis element directly and cannot be classically postprocessed into an alternate basis.
Correspondingly, photon counts can be labeled or tagged by a specific basis in which they trigger.
Finally, the basis is chosen to be orthogonal to the DC fringe offset, which fundamentally changes the nature of quantum noise.

\subsection{Photon Counting}~\label{subs:pc}

Photon counting as a readout scheme is motivated by the fact that quantum noise dominates in certain frequency bands.
In current and next-generation observatories, the classical noise is substantially smaller than the quantum noise frequencies above ${\sim}1\,$kHz, even with squeezing: $S_n(f) \ll S_{\textsc{hd}}(f)$ and $S_{\textsc{hd}}(f) \approx e^{-2z} S_q(f)$. 
In this regime, Fisher Information arguments, App.~\ref{app:FI} and Ref.~\cite{Gardner:2024frf}, show that photon counting may achieve favorable scaling compared to homodyne.
The Cramer-Rao bound for estimating spectral parameters of a stochastic signal scales as ${\propto} \sqrt{S_n(f) S_q(f)}$ under photon counting rather than ${\propto} S_{\textsc{hd}}(f) = S_n(f) + S_q(f)$ under homodyne~\cite{Gardner:2024frf}.
In a stochastic signal, the (complex) GW amplitude in each frequency bin is a random realization drawn from a distribution determined by the signal's spectral density.  
The advantage of photon counting is realized by this randomness, that translates to randomness in the basis coefficients across frequencies.

{The application of BNS postmerger signals is subtly different, as these signals are  stochastic on a per-event basis due to marginalized parameters. The parameter being estimated, $R_{1.6}$ -- a proxy for the EoS, is not random per-event but is unknown over the population. In each measured event, this astrophysical parameter is deterministically mapped to waveform center frequency, which is thus also not expected to be stochastic per event. This BNS frequency-estimation application does not yet have a complete analytic justification for acceleration through photon counting, so our arguments here are heuristic and serve as motivation for the subsequent full calculations.}
For each BNS event, the event-level parameters, such as the postmerger peak frequency or phase, are not random but rather determined by $h_{\textrm{sig}, \theta}$.
However, when combining these event-level parameters to hierarchically compute population-level parameters, such as the NS EoS or the mass distribution, there is an element of randomness due to imperfect knowledge or estimation of $\theta$ on each event.
The exact realization of the observed event is a random draw from the population distribution, be it the NS mass or the signal phase.\footnote{Translating the stochastic signal to the terminology, the ``event-level parameters'' would be the complex GW amplitude in each frequency bin which are hierarchically combined to the ``population-level'' power spectrum.}
Thus, this application reflects a mixture of deterministic and random parameters.
While all BNSs share the same EoS, each signal has a random phase drawn from (presumably) a uniform distribution.
Prior work shows a quantum measurement advantage to photon counting for measuring population-level random parameters, i.e., a population distribution with some width~\cite{Gardner:2024frf}.
This work examines population-deterministic parameters, i.e., BNS radius and EoS, while nuisance parameters that have to be marginalized over, e.g., the phase, are population-level random.

The crucial difference between photon counting and past incremental improvements in the readout of GW detectors --- heterodyne (initial LIGO) to DC readout (advanced LIGO) to (upcoming) balanced homodyne readout (A+ LIGO)~\cite{fritschel2014balanced} --- is that photon counting \emph{fundamentally changes the output measurement data}. 
Rather than recording a continuous representation of the strain present on the detector, instead a discrete set of photon counts are recorded for a discrete set of elements of a temporal mode basis.
The basic operational configuration of the detector remains unaltered, including the fringe offset that enables homodyne readout.
Different readout schemes can be chosen per frequency band, for instance photon counting above $1\,$kHz where quantum noise dominates and standard homodyne readout below this frequency.

The principle behind photon counting is to change from a quantum measurement of the \emph{amplitude of the optical field} in each mode to the \emph{occupation number or energy} in each mode. 
By measuring the number of photons in each temporal mode basis, the underlying signal can be reconstructed. 
Over many events, we can estimate the average occupation per mode and the \emph{RMS magnitude} of the basis coefficients. 
For signals where the coefficients are constant, i.e., fully determined by the signal model given signal parameters, photon counting is not expected to fundamentally accelerate inference.
However, for inference setups where the coefficients have randomness, for instance due to inherent signal stochasticity (such as a truly stochastic signal) or due to population randomness (such as the exact phase and mass of a BNS as drawn from the population distribution), this measurement potentially reduces quantum noise~\cite{Gardner:2024frf}.

Under homodyne, quantum shot noise on the fringe offset photons manifests as an additional Gaussian noise.\footnote{Measuring the field amplitude of an electromagnetic field is akin to measuring the Gaussian wave function of a particle in the position basis, which results in a normal distribution. Squeezing reduces the width of the wave function and improves the statistics, but is limited by state impurity from optical losses.} 
Under photon counting, we select a basis orthogonal to the DC fringe offset, so that the quantum noise is fundamentally different.
The output power effectively reduces to the previously-negligible $\qrv{P}_\epsilon(t)$ term from Eq.~\eqref{eq:Pepsilon}.
This term contains the signal power, $|e_h(t)|^2$; a quantum-state power term,
$\qv{\fE}^\dagger_q(t)\qv{\fE}_q(t)$, that is zero when unsqueezed vacuum states are injected; and a differential-arm shot-noise term, $\qv{\fE}^\dagger_q(t)\rv{\fE}_h(t) {+}  \rv{\fE}^*_h(t)\qv{\fE}_q(t)$. 
This final term is the source of quantum noise in photon counting and amounts to shot noise from photons produced by the signal and classical noise fields, which are small, rather than from the local oscillator.
This term is curious as it injects less and less noise as the signal (and classical noise) becomes diminishingly small, very unlike the homodyne case where shot noise does not depend on $h(t)$ and is instead caused by the fringe light.
The different noise source leads to different statistical behavior.

In other words, the shot noise associated with (potentially squeezed) vacuum fluctuations in homodyne readout does not indicate the fundamental limit on the observability.
In photon counting, it is the rarity of low-SNR signals and the background from classical noise together that provide quantum noise and limit measurements.
The quantum noise (amplitude) spectral density is inversely proportional to the optical calibration $|g(f)|$, Eq.~\eqref{eq:shotnoisePSD}, so the rarity of measuring signals through counts scales with the unsqueezed quantum noise density. 
As a result, Fisher Information for certain kinds of inference has a favorable scaling when the classical noise density is substantially below the quantum noise density.

Formally, for each each temporal-mode $d_k(f)$, we define an operator for the optical field of Eq.~\eqref{eq:e_out_HD}
\begin{equation}
  \qrv{\fE}_k
  = \int_{-\infty}^{\infty}\dd f\, d^*_k(f)\frac{\qrv{\fE}_{\text{out}}(f)}{\sqrt{\hbar k c}}
  = \frac{\big(d_k|\rv{e}_h\big)_{\textsc{tm}}}{\sqrt{\hbar k c}} 
  + \qv{e}_{q,k} \,.~\label{eq:E_k}
\end{equation}
Here $d_k(f)$ has units of $1/\sqrt{\textrm{Hz}}$ while $\rv{e}_h(f)$ has units of $\sqrt{\textrm{J}/\textrm{Hz}}$, Eq.~\eqref{eq:E_h}. Since we divided by the photon energy, $\qrv{\fE}_k$ is unitless.
For temporal modes that are orthogonal to the DC field component, ${\int_{-\infty}^{\infty}\dd f\, d^*_k(f) = 0}$, the fringe term contribution vanishes. 
The term $\qrv{e}_{q,k} = {\big(d_k|\rv{e}_q\big)_{\textsc{tm}}}/{\sqrt{\hbar k c}}$ is the quantum term for the specific $d_k$ temporal mode.

The per-mode operators of Eq.~\eqref{eq:E_k} allow us to compute expectation values, moments, and other statistics for measurements. 
In photon counting, we are interested in the expectation values and moments for measurements of the $k$-indexed photon number, $\qrv{N}_k \equiv \rv{\fE}_k^\dagger \rv{\fE}_k$. 
To simplify notation, we introduce one more inner product that is weighted only by the quantum noise PSD
\begin{align}
\big(d_k|\qrv{x}\big)_{\textsc{pc}} &\equiv \int_{-\infty}^{\infty} \dd f d^*_k(f) \frac{\qrv{x}(f)}{\sqrt{2 S_q(f)}}\,,\label{eq:inner_pc}
\end{align}
yielding through Eqs.~\eqref{eq:E_h} and~\eqref{eq:shotnoisePSD}
\begin{equation}
  \qrv{\fE}_k
  =
\big(d_k|h_{\text{sig};\theta}\big)_{\textsc{pc}}
+
\big(d_k|\rv{n}\big)_{\textsc{pc}}
+
\big(d_k|\qv{e}_q\big)_{\textsc{pc}}\,.~\label{eq:E_k2}
\end{equation}
The expected number of photons in the $k$-th template given an incoming signal sums the signal and classical noise contributions
\begin{align}
  \bar{\pN}_k &= \Braket{\qrv{\fE}_k^\dagger \qrv{\fE}_k }
  =\left|\big(d_k|h_{\text{sig},\theta}\big)_{\textsc{pc}}\right|^2 + \braket{\left|\big(d_k|\qrv{n}\big)_{\textsc{pc}}\right|^2}\nonumber\\
  &=
 \bar{\pN}_{\textrm{sig},k, \theta} + \bar{\pN}_{\textrm{cl},k}\,,
  \label{eq:nk_tot}
\end{align}
where we use the linearity of the inner product with the zero-mean expectation of $\rv{n}$, and the fact that the quantum term corresponds to a vacuum state and thus does not produce photons. 

Experimentally, when an observation is made, a \textit{discrete} number of photons are generated in each mode basis filter. 
The likelihood function is then based on the probability of observing a particular realization of photon count per template.  
For the remainder of this study, we assume perfect efficiency in the photon readout devices.

Finally, we summarize the key formal difference in the quantum noise between the two readout schemes. 
Counting photons contain the quantum operator term $\qrv{e}_{q,k}^\dagger\qrv{e}_{q,k}$, which is a number operator for photons in the temporal mode $d_k$. 
This quantum term creates the specific statistical distributions detailed below. 
By contrast, the quantum noise term for homodyne is $\qv{P}_q = e_{\text{fringe}}\qv{e}_{q}(t)^\dagger + e^*_{\text{fringe}}\qv{e}_{q}(t) \propto \qv{q}$. This term is known as a quadrature field operator, has a continuous distribution, and imparts Gaussian statistics to the measurement.

\subsubsection{Classical noise photon distribution}

Classical noise contributes a \textit{background} of photons with expected number
\begin{align}
  \bar{\pN}_{\textrm{cl},k} &= \braket{|\big(d_k|\qv{n}\big)_{\textsc{pc}}|^2} \nonumber
  \\ &=
       \int_{-\infty}^{\infty}\dd f\int_{-\infty}^{\infty}\dd f'\, d_k(f) d^*_k(f')\frac{\braket{\rv{n}^*(f)\rv{n}(f')}}{2S_q(f)}\,.
       \nonumber
    \\ &=
    \int_{-\infty}^{\infty} \dd f\, |d_k(f)|^2 \frac{\frac{1}{2}S_n(f)}{2S_q(f)}\approx \frac{S_n(f_k)}{4S_q(f_k)}\,,~\label{eq:nbarcl}
\end{align}
where the approximation holds for templates that are well-concentrated around each of their center frequencies, Eq.~\eqref{eq:template-fk}, compared to the variation in $S_q(f)$ and $S_n(f)$.
Overall, the number of background photons scales with the classical noise PSD.

The distribution from which this count is generated depends on the noise sources as a result of different occupation of the state in phase space. 
For the majority of noise sources, and all noises that we will consider here, classical noises purely manifest in a single quadrature of the light, i.e., $\rv{n}(t)$ is a real number that is random. 
Classical randomness into a single optical field quadrature leads to a geometric-like distribution\footnote{{This is Eq.~(12) in \cite{Gardner:2024frf} with the conversion of quadrature-operator variance, $\sigma^2$ (their notation) to average number, $\bar{\pN}_{\textrm{cl},k}$ (our notation) of $\sigma^2 = 2\bar{\pN}_{\textrm{cl},k}$. This relation follows from the vacuum having 1/2 variance in either quadrature (their notation) while having a total zero-point energy of 1/2 a photon (1/4 photon from each quadrature, our notation).}} with mean $\bar{\pN}$ and variance $\bar{\pN} + \bar{\pN}^2$~\cite{Gardner:2024frf}.
\begin{equation}
\mathcal{P}_{\text{cl}}(\pN_{k}|\bar{\pN}_{\textrm{cl},k}) = \frac{(2\pN_{k})!}{2^{\pN_{k}}(\pN_{k}!)^2}\frac{(\bar{\pN}_{\textrm{cl},k})^{\pN_{k}}}{(2\bar{\pN}_{\textrm{cl},k}+1)^{\pN_{k}+1/2}}\,.~\label{eq:gardner}
\end{equation}

\subsubsection{Signal photons}

The expected number of signal photons is
\begin{align}
  \bar{\pN}_{\textrm{sig},k, \theta} &= \left|
    \int_{-\infty}^{\infty}\dd f\, d^*_k(f)  \frac{h_{\textrm{sig}, \theta}(f)}{\sqrt{2S_q(f)}}\right|^2\approx
  \Delta f_k \frac{|h_{\textrm{sig}, \theta}(f_k)|^2}{2S_q(f_k)}\,,~\label{eq:nbarsig}
\end{align}
where the approximation holds for signal power and quantum noise that are relatively constant and the template covers a bandwidth
 $\Delta f_k$, Eq.~\eqref{eq:template-dfk}, and has a central frequency $f_k$, Eq.~\eqref{eq:template-fk}.
 The amount of signal photons scales quadratically with the signal amplitude.
Using a complete and orthonormal basis, one can show
 \begin{align}\label{eq:PC_total_power}
   \sum_k \bar{\pN}_{\textrm{sig},k, \theta} = \int_{-\infty}^{\infty}\dd f\, \frac{|h_{\textrm{sig}, \theta}(f)|^2}{2S_q(f)}\,.
 \end{align}
The total number of photons emitted by a signal is a factor of 4 lower than the standard homodyne optimal SNR squared of Eq.~\eqref{eq:snr} when the shot noise PSD alone is used in the denominator.

For each signal, the output field selected by the basis is a coherent state. 
Formally, this occurs because $\big(d_k | h_{\text{sig}, \theta}\big)$ in Eq.~\eqref{eq:E_k} is a complex-number and not an operator. 
The number of detected signal photons then follows a Poisson distribution with mean $\bar{\pN}$ and variance $\bar{\pN}$,
\begin{equation}
    {\cal P}_{\text{sig}}\Big(\pN_{k}\Big|\bar{\pN}_{\textrm{sig},k, \theta}\Big) = e^{-\bar{\pN}_{\textrm{sig},k, \theta}}\frac{(\bar{\pN}_{\textrm{sig},k, \theta})^{\pN_{k}}}{\pN_{k}!}\,.~\label{eq:poisson}
\end{equation}

\subsection{Photon counting likelihood for transient signals}~\label{subs:pc_like}

During a measurement, a total number of photons $N_k$ is measured in each temporal mode, which is a combination of signal photons and the classical noise background.
The likelihood for an observation of $\pN_k$ photons from a given basis is then a convolution, marginalizing over whether each photon is associated with a transient signal or classical noise processes
\begin{equation}
    p(\pN_k | \theta) = \sum_{n=0}^{\pN_k}{\cal P}_{\text{sig}}(n|\bar{\pN}_{\textrm{sig},k,\theta})\mathcal{P}_{\text{cl}}(\pN_k - n |\bar{\pN}_{\textrm{cl},k})\,.\label{eq:likelihood_indiv}
\end{equation}
Above, we rely on a measurement of the classical noise PSD to calculate $\bar{\pN}_{\textrm{cl},k}$ using Eq.~\eqref{eq:nbarcl}. 
As $\bar{\pN}_{\textrm{cl},k}\to0$, Eq.~\eqref{eq:likelihood_indiv} reduces to the Poisson distribution.
Intuitively, in the single photon limit, this expression reduces to the sum of the probabilities that the photon was a signal photon multiplied by the probability that the photon does not originate from a noise process and vice versa (signal photon and not a noise photon). 
While the probabilities are fixed for the classical noise photons (under the assumption that the PSD is known), varying the parameters of the signal $\theta$ will lead to different probabilities of $\mathcal{P}(n|\bar{\pN}_{\textrm{sig},k,\theta})$, thereby enabling inference of the model parameters. 

The full likelihood for a photon counting readout over $M$ orthonormal filters is the product of the individual likelihoods from each filter
\begin{equation}
    p(\{\pN_k\}|\theta) = \prod_{k=0}^M p(\pN_k | \theta)\,.\label{eq:likelihood}
\end{equation}
This is analogous to the Whittle likelihood for homodyne readout, where the total likelihood is the product of the likelihoods per individual frequency bin~\cite{vanderSluys:2008qx,Veitch:2008ur}.
With the likelihood of Eq.~\eqref{eq:likelihood} and a corresponding prior for the signal parameters $\theta$, we can proceed with traditional GW inference under a photon counting readout. 

\subsection{Implementation Requirements}
With the interferometer and readout formalism detailed above, we now discuss the principal requirements for any future implementation of temporal-mode photon counting in an interferometer observatory such as CE or LIGO.

First, the primary requirement of the readout is the temporal mode selectivity. 
The ability to tag photons in a chosen temporal mode basis is physically allowed and enables parameter estimation. 
Second, efficiency depends on the background ``dark'' count rate of the photon-counting detectors. 
Equation~\eqref{eq:nbarcl} establishes the classical noise rate for any single temporal mode in the detector. 
Figure~\ref{fig:noise_scaling} (detailed later) shows the noise performance of CE in the relevant detection bands, with an unsqueezed shot-noise PSD of $10^{-49}/\text{Hz}$ and classical noise at  $10^{-51}/\text{Hz}$. 
This indicates that the classical noise is equivalent to one dark count per 200 events. 
The detectors then need a performance that is a factor of ten better than this, one per 2000 events, for their per-event background (false alarm) count rate.
This background count rate will be challenging for future hardware due to the leakage of mW-levels of light from the interferometer. 
The related requirements are well-beyond the scope of this work, but being explored in ongoing experiments~\cite{Vermeulen:2024vgl}.

Furthermore, there is a potential trade-off with the use of squeezing, which carries energy to create background photon counts. 
Squeezed states of light can be manipulated and injected in a time and frequency-dependent manner, as demonstrated in LIGO~\cite{Ganapathy:2023}. 
Any future implementation of photon counting would prevent squeezed states from contaminating the injected vacuum states assumed in this work, either by temporally gating them, using similar technology as the readout to selectively absorb them, or by using cavities to filter them away. 
In summary, there is no fundamental incompatibility of the planned use of squeezed states and the introduction of photon-counting technology. 
Furthermore, Ref.~\cite{Gardner:2024frf} showed that quantum state enhancement can also enhance the readout studied here, but we do not analyze this possibility in this study.

Finally, there is the impact of readout inefficiency. 
Unlike squeezing, loss or inefficiency in photon counting only reduces the effective optical gain $g(f)$ (see Eq.~\eqref{eq:cal_gain}).

\section{Individual-event postmerger inference}~\label{sec:single}

Having laid out the the photon counting formalism and its relation to homodyne in Sec.~\ref{sec:pc}, we apply it to BNS postmerger signals. 
Since the signals are anticipated to be between $1.5$ and $4\,$kHz where quantum shot noise dominates, $S_q(f) \gg S_n(f)$, 
they are an ideal example of photon counting's application.
Here we beginning with exploring inference for a single event, outlining the basis filters in Sec.~\ref{subs:design} and exploring the impact of noise background, photon counts, and SNR in Sec.~\ref{subs:highlow}.

\subsection{Filter design for postmerger signals}
\label{subs:design}

The temporal mode filter response needs to be designed to closely mimic the targeted signal, specifically postmerger signals.
Figuratively, this can be understood as in-situ hardware-based matched-filtering where the basis must closely match the signal, i.e., maximize Eq.~\eqref{eq:nbarsig}. 

postmerger signals are dominated by peaks in the frequency spectrum which can be modeled as damped sinusoids. 
The general function for the damped sinusoid in the frequency domain is 
\begin{align}
  L(f | f_0, \gamma, A, \phi_0, t_0)
  &=
                                       \nonumber\\&\hspace{-3.5em}A\gamma  \Bigg(\frac{e^{-i(\phi_0 - 2\pi t_0 f_0)}}{\gamma + i(f_0-f)} + \frac{e^{i(\phi_0 - 2\pi t_0 f_0)}}{\gamma + i(f_0+f)}\Bigg)\,,~\label{eq:basis}
\end{align}
where $f_0$ is the peak frequency, $\gamma$ is the half-width at half-maximum, $A$ is the amplitude (in the time domain), and $\phi_0$ and $t_0$ are phase and time offsets.
The amplitude profile of Eq.~\eqref{eq:basis} is a Lorentzian.
 
Given a signal with duration $\Delta T$ and bandwidth $\Delta F$, the number of filters required is 
\begin{equation}
    N \sim 2 \Delta F \Delta T\,.
\end{equation}
This follows directly from the number of terms in the signal Fourier transform, with the factor of two accounting for both sine and cosine components. 
{For redshifted postmerger signals, $\Delta F\approx 3.0\,\textrm{kHz}$ and $\Delta T\approx 50$\,ms, resulting in $300$ filters (150 sine and 150 cosine filters). }
We construct filters with the following parameters
\begin{enumerate}
    \item $A$ is set by orthonormalization (discussed shortly); 
    \item $\gamma$ is set to $20\,\textrm{Hz}$ (FWHM, nonangular), slightly narrower width than the postmerger peak as modeled by Ref.~\cite{Soultanis:2024pwb} {for redshift $z=1$ events}; 
    \item $f_0$ spans $0.6$ to $4\,\textrm{kHz}$ with 40 {log-spaced} points;
    \item $t_0$ spans $-20$ to $30$\,ms with 5 equally spaced points;
    \item $\phi_0$ is set to either $0$ or $\pi/2$ for the cosine or sine components. 
\end{enumerate} 
Upon construction, the damped-sinusoids are shuffled into a random order and the {orthogonalization (i.e. Gram Schmidt or QR decomposition)} method is applied to obtain a set of orthonormal filters, $\{d_k(f)\}$. 
The resulting {400} filters may not be optimal since the basis of Eq.~\eqref{eq:basis} does not exactly align with all postmerger signals. The {slightly overcomplete} construction given above captures around {80\%} of the total signal energy of most signals (in contrast to the expectation from Eq.~\eqref{eq:PC_total_power}).

\begin{figure*}
    \centering
    \includegraphics[width=1.0\linewidth]{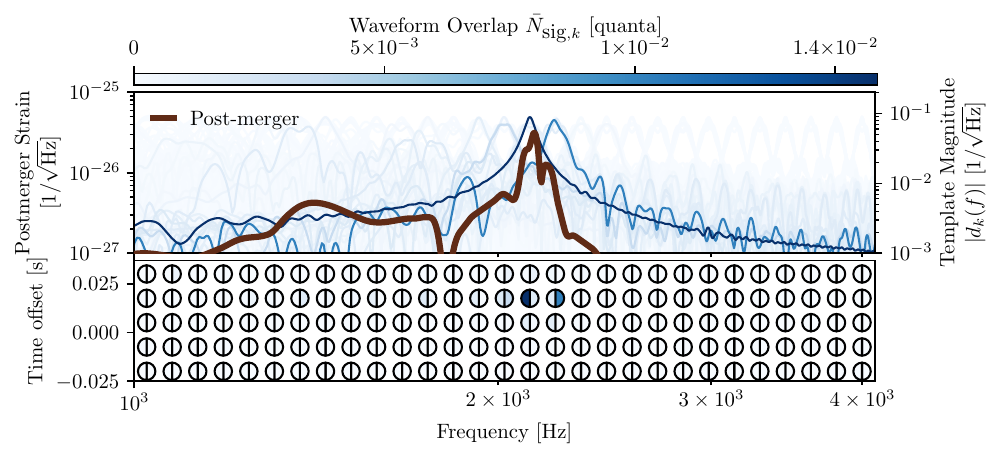}
    \caption{
    Demonstration of the detection of a BNS postmerger signal with a damped-sinusoid temporal mode basis. 
    In the upper panel, we show the strain (left vertical axis) and basis filter (right vertical axis) spectral amplitudes as a function of frequency. 
    The basis modes are colored according to their expected number of signal photons. 
    While each basis mode is initially constructed according to Eq.~\eqref{eq:basis}, orthonormalization leads to more complex basis filter structure. 
    In the lower panel, we present the 200 basis filters in terms of their time, frequency, and phase (sine on the left half circle and cosine of the right), again colored by the expected number of signal photons. }
    \label{fig:filters}
\end{figure*}

As an illustration of the filters in action, we simulate a postmerger signal with a supervised learning model trained on numerical relativity simulations~\cite{Soultanis:2024pwb} under the APR4 EoS. 
The signal has a total mass of $2.4\,M_\odot$ and redshift of $0.05$, and has an optimal SNR of 1.10 in an unsqueezed version of CE. 
The postmerger peak frequency is $\sim3100$\,Hz. 
We plot the signal and the temporal mode basis functions colored by their expected number of signal photons in Fig.~\ref{fig:filters}. 
In the upper panel, we show the signal and filter spectral amplitudes. 
The basis filters possess nonregular patterns and features as a result of the orthonormalization procedure, but are characterized by broad spectral peaks. 
In the lower panel, we represent the time, frequency, and phase of the basis. 
The filters will not exactly correspond to these grid points due to the orthonormalization. 
However, it is a useful visual indicator of which bases can result in a photon generation. 
Unsurprisingly, the greatest number of expected photons originate from the templates which overlap the most with the signal. 

\subsection{Single-event inference of damped sinusoids}
\label{subs:highlow}

Given the photon counting likelihood in Eq.~\eqref{eq:likelihood} and the filters of Sec.~\ref{subs:design}, we simulate postmerger signals according to Eq.~\eqref{eq:basis}, and obtain the posterior distribution under flat priors on all parameters $\theta = \{f_0, \gamma, A, \phi_0, t_0\}$. 
The peak frequency ranges from $0.1$ to $4$\,kHz, $\gamma$ ranges from $0$ to $100$\,Hz, $A$ from $0$ to $10$-times the expected magnitude. $\phi_0$ {ranges from $-2\pi$ to $2\pi$ to reduce sampling artifacts near phase 0 on the uniform distribution and wrapped to $0, 2\pi$ in the figures.} {The ringdown start-time $t_0$ ranges from -20\,ms to +20\,ms to account for uncertainty in the start of the postmerger}.
While for illustration we have chosen the same functional form for the basis (prior to orthonormalization) and the signal, these do not need to be same.  
We consider different scenarios and compare to an unsqueezed version of CE and homodyne readout. 
In the following illustrative examples, we consider no noise contamination due to Gaussian noise fluctuations for either the homodyne or the photon count readout analyses. 

\begin{figure*}
    \centering
    \includegraphics[width=\linewidth]{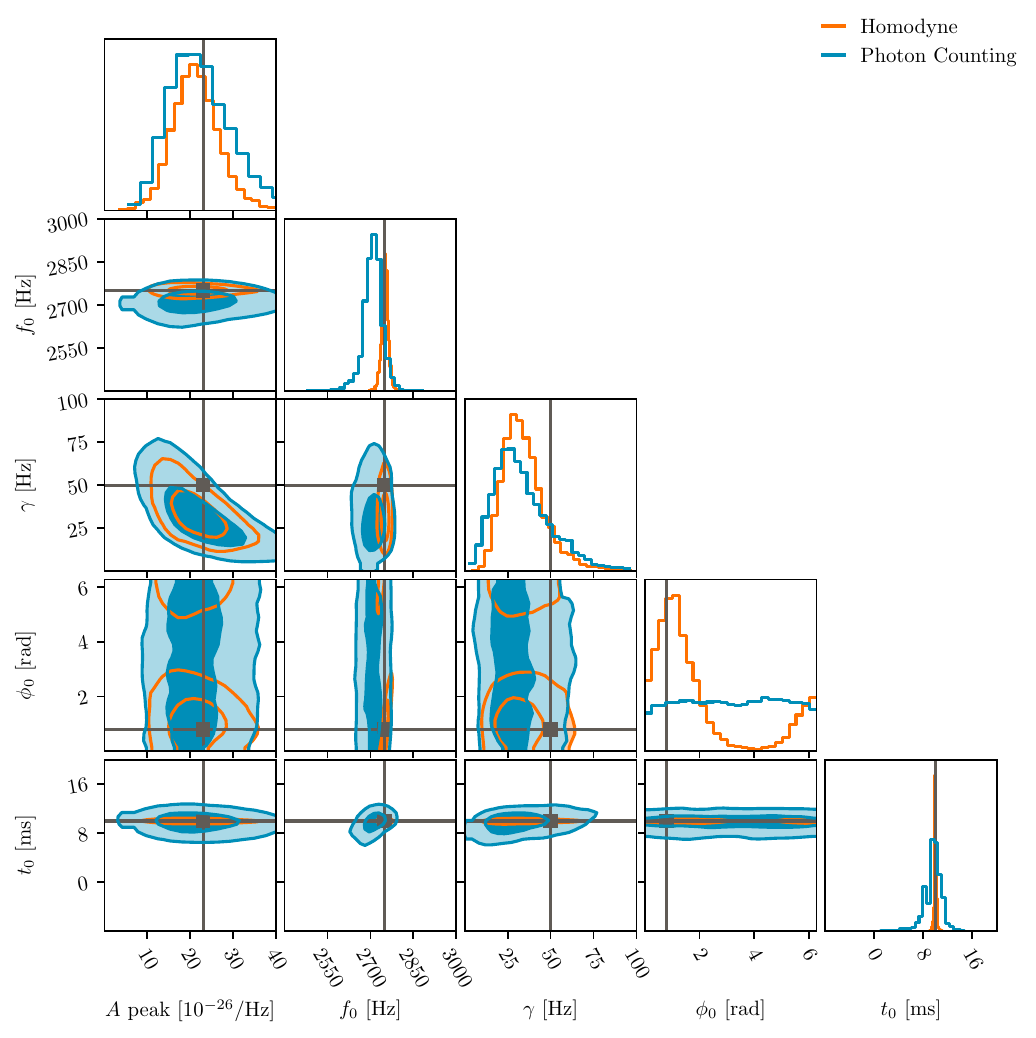}
    \caption{Marginal posterior distribution for the damped-sinusoid parameters obtained with the homodyne (orange) and photon counting (blue) readouts for a signal with SNR 5. 
    The true values are shown with grey lines. 
    Contours correspond to the 50\% and 90\% credible levels. 
    Overall, homodyne constraints are more stringent in this relatively high-SNR regime --- to be expected.
    The time and phase posterior distributions are bimodal with the photon counting readout due to two different temporal mode filters, $\{d_k\}$, generating photons.}
    \label{fig:corner_super}
\end{figure*}

We begin with a super-threshold observation --- a postmerger signal that would be detected with a standard homodyne readout. 
We simulate a damped-sinusoid signal with an optimal SNR of 5, a peak frequency of $2.75\,$kHz, $\gamma=50$\,Hz, $\phi_0 = \pi/4$, and $t_0 = 10$\,ms. 
The expected number of signal photons is $6.25${, with our template grid only overlapping with $4.6$ photons worth of power (worse than its average efficiency)}.
The number of detected photons was $3$.
In Fig.~\ref{fig:corner_super}, we show that both readouts can confidently constrain all parameters of such a loud signal. 
The homodyne generally outperforms photon counting unless counting gets above its mean number of photons, constraining the peak frequency 3 times better at the 68\% level.
Since at least one photon was detected, the photon counting amplitude definitively rules out zero.
The homodyne amplitude is consistent with zero, albeit constrained 1.5 times better. 

\begin{figure*}
    \centering
    \includegraphics[width=\linewidth]{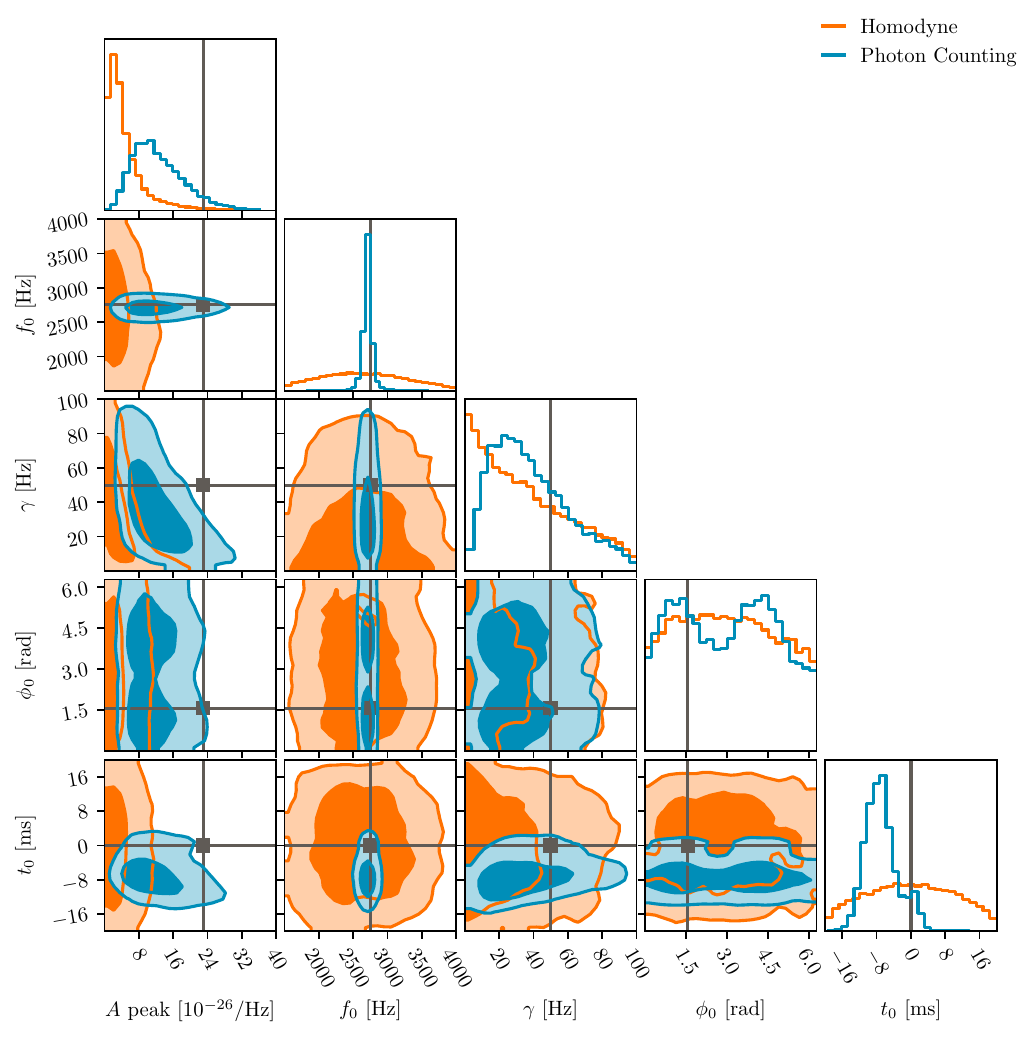}
    \caption{Same as Fig.~\ref{fig:corner_super} but for a signal with SNR 1. 
    Such a low SNR signal in the homodyne readout {does not constrain the parameters}. 
    In the case of photon counting, $\bar{\pN}_\textrm{sig} = 0.125$, and so there is a $11.8\%$ chance that at least one photon is generated. 
    If a photon is recorded, the signal is detected and its parameters are constrained.}
    \label{fig:corner_sub}
\end{figure*}

We now turn to a more realistic example, that of a subthreshold observation of an SNR 1 signal --- the same damped sinusoid with a smaller amplitude. 
The expected number of photons given the ${\sim}80\%$ efficiency of the filter bank and available photons is $\bar{\pN}_\textrm{sig} = 0.2$, corresponding to a $18\%$ chance that at least a single photon is generated. 
Figure~\ref{fig:corner_sub} shows results for the case where a photon is indeed generated and recorded, demonstrating the expected benefit of the photon counting readout scheme. 
While the homodyne readout recovers no information from such a low-SNR observation, the photon counting readout can detect the signal and constrain the peak frequency and time offset. 
Such constraints would be expected for about 1 in 5 signals with SNR 1.

\subsection{Impact of noise backgrounds and photon counts}

Photon counting inference has two {notably} un-intuitive elements.
Unlike homodyne where the posterior width is only related to the inverse of the total PSD $S_\textrm{HD}(f)$, the photon counting posterior depends on \textit{both} the relative relation between the signal to noise (quantum and classical), \textit{and} the number of signal photons generated which is an inherently random process. Here we elaborate upon both. 

\begin{figure*}
    \centering
    \includegraphics[width=\linewidth]{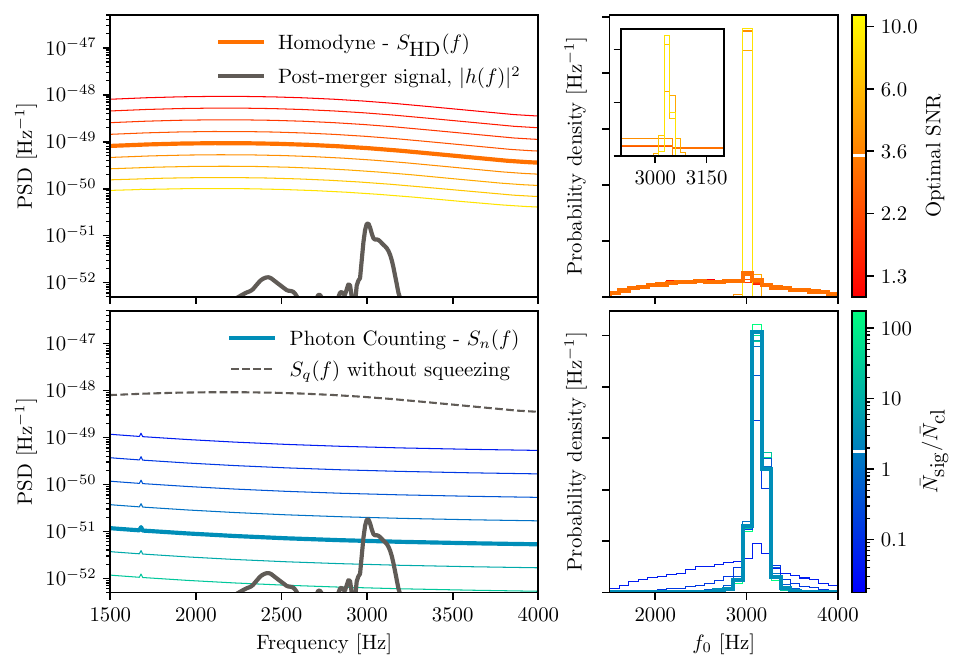}
    \caption{Impact of the noise backgrounds for the homodyne (upper; orange) and photon counting (lower; blue) readouts. 
    The left panel shows the postmerger signal spectrum, as well as $S_\textrm{HD}(f)$ for the homodyne and $S_n(f)$ for the photon counting at different squeezing (top) and classical noise (bottom) levels. The squeezing and classical noise are scaled by $\pm 10\text{dB}$ from their nominal levels.
    The relevant statistic for the homodyne readout is SNR $\sim 1/S_\textrm{HD}(f)$, while the relevant statistic for the photon counting is $\bar{\pN}_\textrm{sig}/\bar{\pN}_\textrm{cl} \sim |h(f)|^2/S_n(f)$; both are denoted in the colormap.
    The thicker lines correspond to the expected results with CE's design squeezing level (10\,dB; for the homodyne), or classical noise realization (for the photon counting), and the white lines on the colorbars indicate their corresponding values. In the right panel, we show the marginal posterior for the peak frequency for each corresponding noise level. 
    }
    \label{fig:noise_scaling}
\end{figure*}

Regarding the noise sources, Eqs.~\eqref{eq:nbarcl} and~\eqref{eq:nbarsig} imply that the expected number of photons for classical noise and signal are $\bar{\pN}_\textrm{cl} \sim S_n(f)/S_q(f)$ and $\bar{\pN}_\textrm{sig}\sim\abs{h(f)}^2/S_q(f)$ respectively. 
With these in mind, we have the following possibilities. 
\begin{enumerate}
    \item The high SNR regime, where the signal is comfortably above the noise backgrounds, is $\abs{h(f)}^2 \gg S_{q}(f)$ \& $ S_n(f)$. In this case, the relative sensitivity loss due to quantum effects and classical noises is unimportant. This is analogous to Fig.~\ref{fig:corner_super} where a homodyne readout is expected to outperform photon counting {due to the SNR increase from squeezing (otherwise they are similar)}. 
    \item Classical noise is higher than the signal, $S_n(f) \gg \abs{h(f)}^2$. The homodyne readout then suffers from significant Gaussian noise due to classical effects. 
    The photon counting readout will record a high number of noise photons, making it hard to pick out the signal photons. 
    If $S_q(f) \gg S_n(f)$, each observation results in few photons, but the classical noise ones still dominate over an ensemble. 
    No readout choice will lead an observable signal. 
    Conversely, if $S_n(f) \gg S_q(f)$, each observation results in many photons. 
    \item The low-SNR regime, $S_q(f) \gg \abs{h(f)}^2 \gg S_n(f)$, will produce more signal than noise photons. This is analogous to Fig.~\ref{fig:corner_sub}. 
    Here photon counting shows the most promise as statistically after many observations, some photons will be recorded. For LIGO, this noise hierarchy corresponds to the frequency band above $\sim1$\,kHz. 
\end{enumerate}

We explore these cases in Fig.~\ref{fig:noise_scaling} with the simulated signal from Fig.~\ref{fig:filters} at an optimal SNR of 1.10 in an unsqueezed CE. 
Photon counting has an expected photon count rate of {0.24}, and therefore a {19\%} chance to generate a single photon and a 2.2\% chance to generate two photons. 
We vary the overall sensitivity by scaling the shot noise in the homodyne readout (top) or the classical noise background in the photon counting readout (bottom). 
The homodyne range is approximately equivalent to dB of squeezing ranging from 0 to 20\,dB.
The thicker lines correspond to the expected total noise for CE~\cite{Evans:2021gyd} (including squeezing; $\sim$ 10\,dB) for the homodyne results, and to the expected classical noise background for the photon counting results. 
In the left panels we show the spectral power for the signal and noise and in the right panels we show the marginal posterior for the peak frequency. 
In all cases, a single photon is detected in the most likely basis filter following Eq.~\eqref{eq:nbarsig}. 

\begin{figure*}
    \centering
    \includegraphics[width=\linewidth]{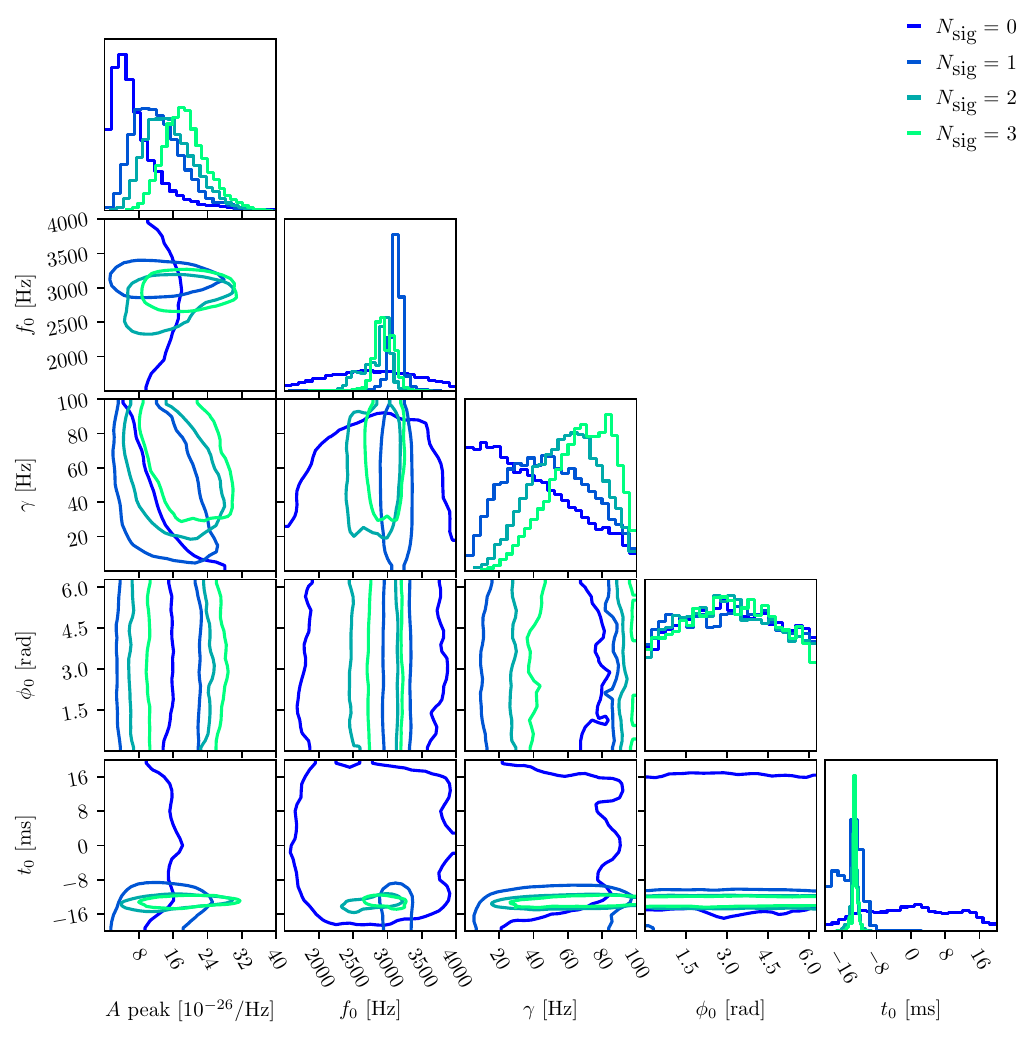}
    \caption{Marginal posterior (90\% credible levels) for  simulated postmerger signals with varying photon counts. 
    As the photon count increases, constraints narrow in a manner similar to homodyne readout. 
    Each case corresponds to a specific realization of a simulated signal resulting in the indicated numbers of photons for illustration. At N=1, the frequency is resolved to be narrower since only one bin has a count. At N$>1$, bins at different frequencies were triggered, spreading the most likely signal to one that overlaps both bins.
    For the SNRs anticipated for next-generation detectors, $\pN_\textrm{sig} > 1$ will be rare.
    }
    \label{fig:n_photon}
\end{figure*}

The two readouts result in fundamentally different behaviors for the posterior as the noise level changes. 
In a homodyne readout, as the SNR increases, the overall width of the posterior decreases continually. 
However, this is only true up to a point for photon counting. 
Once $\bar{\pN}_\textrm{sig} \gtrsim \bar{\pN}_\textrm{cl}$ (the third case above), the posterior width is unchanging even as the classical noise continues decreasing.
The posterior asymptotes to the resolution attainable from a single photon. 
Once the SNR exceeds $\sim5$, the homodyne readout outperforms photon counting.

Tighter constraints with photon counting are possible if more photons are observed. 
We explore this in Fig.~\ref{fig:n_photon}, with 4 simulations with an increasing numbers of photons up to $3$. 
More photons are in principle possible, but observations would suffer from diminishing returns due to basis misspecification.
For $\pN_\textrm{sig} \geq 2$, the photons are generated by different bases elements. 
An observation has a \textit{total} Poisson probability ${\cal{P}}(\pN_\textrm{sig}|\bar{\pN}_\textrm{sig})$ of generating $\pN_\textrm{sig}$ photons. 
Therefore, there are two ways to generate a given photon count; either increase the signal amplitude such that $\bar{\pN}_\textrm{sig}$ increases, or simply wait until a fortuitous photon count realization occurs. 
Here, as the photon count increases, constraints on all parameters shrink in a similar manner to homodyne constraints under increasing SNR, e.g., Fig.~\ref{fig:noise_scaling}.

To summarize, photon counting readout can improve constraints in the low-SNR regime if, and only if, quantum noise dominates both the signal strain and the classical noise. 
Rather than having a single noise level such as in homodyne readout, the output of photon counting depends on both the quantum noise $S_q(f)$, and the classical noise $S_n(f)$. 
The floor in the posterior distribution is determined by the ratio of the signal strength to the classical noise, $|h(f)|^2/S_n(f)$, while the posterior width is given by the ratio of the signal strength to the quantum noise, $|h(f)|^2/S_q(f)$. 

\subsection{Summary of individual constraints}

\begin{figure*}
    \centering
    \includegraphics[width=\linewidth]{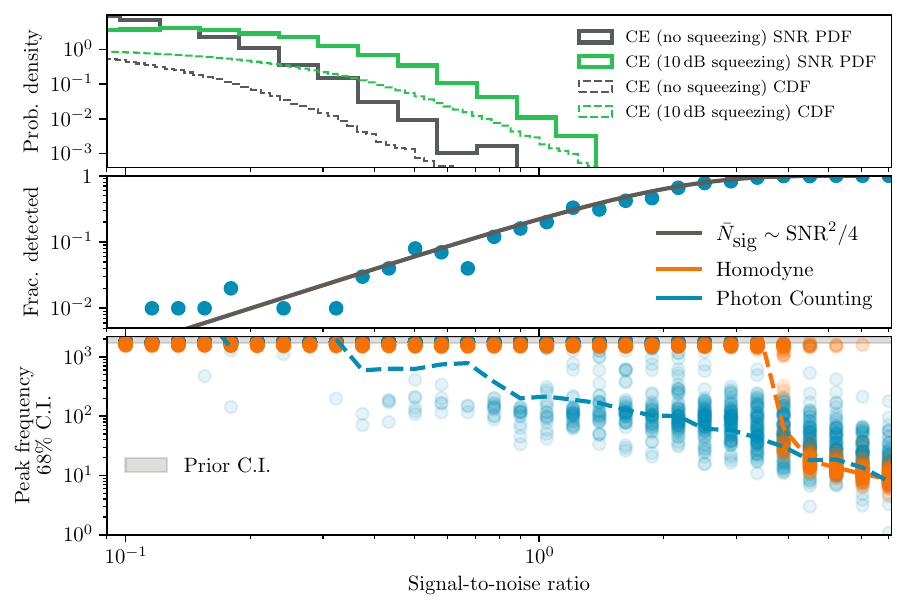}
    \caption{Comparison of photon counting and homodyne as a function of the signal SNR. 
    For reference, the top panel shows the expected distributions of SNRs from $10^4$ observations in an unsqueezed CE (dark grey), as well as at design sensitivity (green).
    The middle panel shows the fraction signals with at least one detected photon, as well as the theoretical expectation if $\bar{\pN}_\textrm{sig}$ follows $\textrm{SNR}^2/4$.
    The bottom panel shows the width of the 68\% credible interval for the peak frequency posterior for each of the 100 simulated events per SNR value between 0.1 and 100 (circles) for homodyne (orange) and photon counting (blue).
    The dashed lines correspond to the harmonic mean of the C.I. over simulated signals. The harmonic mean is used since the C.I. is a proxy for the ``information''}
    \label{fig:snr_summary}
\end{figure*}

We finally consider ensembles of realistic BNS postmerger signals. 
For SNRs ranging from 0.1 to 10, we generate 100 signals per SNR value with the APR4 EoS model~\cite{Soultanis:2024pwb} with uniformly distributed individual masses between $1.2$ and $1.4$\,$M_\odot$. 
We simulate detection with a homodyne readout (CE with no squeezing) and photon counting with no classical noise.
In the bottom panel of Fig.~\ref{fig:snr_summary}, we show the width of the 68\% credible interval of the peak frequency posterior for each simulated signal under both readout schemes as a function of the SNR. 
While homodyne outperforms photon counting for SNRs above ${\sim}5$, i.e., in the individually-detectable signal regime, photon counting leads to on-average better constraints for lower SNRs.
Occasionally, photon counting can detect very low SNR signals. 
Specifically, $\sim1$ in $\mathcal{O}(100)$ observations of an SNR 0.2 signal lead to a photon detection and a measurement of the peak frequency to within $\lesssim 500$\,Hz at the 68\% level. 
In the middle panel, we show the fraction of signals that were detected, and how this relates to the theoretical result when $\bar{\pN}_\textrm{sig} \sim \textrm{SNR}^2/4$.
The detected fraction is typically below the theoretical curve, due to the mismatch between the signal and templates bases --- though low number statistics lead to fluctuations in the lower SNR regime.  
Finally, in the top panel, we show the distribution of expected BNS postmerger SNRs in CE, under no squeezing (relevant for photon counting), or with 10\,dB of squeezing (design sensitivity).
The vast majority of signals will fall in the very low SNR regime where photon counting can provide constraints due to the serendipitous detection of the occasional photon.

\section{Combined Equation-of-state constraints}~\label{sec:hier}

Since most signal will have low SNR, the most stringent constraints will come from combining information across multiple observations.  
To demonstrate the impact of the readout scheme on such hierarchical constraints, we follow \citet{Criswell:2022ewn} and relate the postmerger peak frequency to the binary chirp mass and the radius of a $1.6\,M_\odot$ NS, denoted $R_{1.6}$:
\begin{align}
    f_\textrm{peak} = \beta_0 &+ \beta_1 \mathcal{M} + \beta_2 \mathcal{M}^2 +\beta_3 R_{1.6}\mathcal{M} \nonumber\\&+\beta_4 R_{1.6}\mathcal{M}^2 +\beta_5 R_{1.6}^2\mathcal{M} + \epsilon,~\label{eq:fpeak_eos}
\end{align}
where $\beta_i$ are fit parameters to numerical simulations, given in Table I of Ref.~\cite{Criswell:2022ewn}. 
The additional error term $\epsilon$ is taken to follow a zero-mean Gaussian distribution with a standard deviation of 61\,Hz.   

Again following \citet{Criswell:2022ewn}, we construct the population likelihood for the hierarchical analysis as 
\begin{equation}
    p(\{d\}| R_{1.6}) = \prod_{i=1}^{\pN_\textrm{events}} p\Bigg(d_i\Bigg|f_{0,i} =\frac{f_\textrm{peak}(M_i, R_{1.6})}{1+z_i}\Bigg)\,, ~\label{eq:hier_l}
\end{equation}
where we assume that we know the total mass and redshift from the BNS inspiral portion of the waveform and restrict to equal-mass binaries. 
The peak frequency is then a function of the total mass of the NS binary, and the EoS-dependent radius at 1.6\,$M_\odot$, $R_{1.6}$. 
While this is only approximate, it provides a baseline with which to compare photon counting and homodyne.

Unbiased recovery of the NS radius with Eq.~\eqref{eq:hier_l} hinges on ensuring that the population distribution of BNSs matches the parameter priors used during inference for each individual event.
This can be ensured by generating and recovering the signal with the same distributions. 
In practice, however, since we are generating the postmerger signals with the model of Ref.~\cite{Soultanis:2024pwb}, we do not have immediate access to the implied distributions for the damped-sinusoid parameters. 
To approximate these distributions, we generate $10^4$ signals and extract the best-fit (maximum likelihood) amplitude and $\gamma$ from which we construct the ``astrophysical'' distribution on the damped-sinusoid parameters.  
The primary and secondary binary masses are drawn from a uniform distribution ranging from 1.2 to 1.4\,$M_\odot$, the redshift is drawn from the Madau-Dickinson star-formation rate~\cite{Madau:2014bja}, the sky location is isotropic across the sky, and rotation angles are uniform.  

The simulated population consists of $10^4$ signals from the above distributions.
Due to the uncertainty in the BNS merger rate density~\cite{gwtc3_pop,LIGOScientific:2025pvj}, this corresponds to anywhere between $5\times10^{-3}$ and $0.75$ years of observation.
Under the design sensitivity of CE (with squeezing), 95\% of the signals have SNRs $\lesssim 0.47$, and only {22} signal photons are recorded.
In contrast, {3607} classical noise photons are recorded with the design sensitivity classical noise, down to {388} if the classical noise is reduced by an order of magnitude.
Inference {accounts for the noise background} and {detection hinges on our ability to disentangle} the signal from the noise photons. {In other words, what is important is how the noise and signal photons enter the likelihood and if the signal photons are statistically resolvable. The Bayesian inference presented here shows that the few signal photons ``stack'' sufficiently to constrain the posterior in spite of the background photons, which appear randomly and simply reduce the statistical efficiency of inferences on the signal photons. In contrast, homodyne readout can be thought (heuristically) to have 1/4 of a photon worth of random amplitude in every template. Squeezing reduces this background corresponding the variance reduction in dBs. CE's 8db of squeezing at high frequencies then corresponds to 0.04 photons/temporal-mode, or about 16/per-event -- considerably more total than by photon counting. Bayesian inference on the homodyne readout is known to handle a Gaussian noise background and here we study the ability to handle temporal-mode tagged photon counting.}

The damped-sinusoid parameters are drawn from their population distributions {with the exception of the amplitude parameter. We find that the population distribution of amplitudes biases the hierarchical inference due to the abundance of distant, low-amplitude events. We instead draw amplitudes from a uniform distribution with its maximum value at 10x the actual peak amplitude. We acknowledge a potential for bias from this ``weakly informed'' prior, but choose it for two reasons. First, the uniform prior appears to prevent bias in the final radius estimation, and we see no reason why it should favor either readout scheme. Second, our posteriors show low support at the high-end of this range for either readout, and choosing a low upper limit improves the Monte Carlo sampling efficiency substantially. An alternative choice we have tried is to use the upper limit of a uniform distribution from the loudest sampled event. This produces very similar results but has lower sampling efficiency.}

{As mentioned in our overview and justification, marginalizing over damped-sinusoid parameters per Sec.~\ref{subs:highlow} affects the resolving ability of the inference. Specifically, we impose a uniform prior on the phase parameter and the start time, which is motivated from an expected inability to precisely resolve the time of merger for the majority of events. In App.~\ref{app:hierarchicalB} we show results that assume perfect knowledge of the merger time for comparison.}

The error term $\epsilon$ is randomly drawn from its appropriate distribution.
Ultimately, this is an idealized inference scenario, but still enables a comparison between the two readout schemes. 

For each choice of $R_{1.6}$, we compute $p(\{d\}|R_{1.6})$ in Eq.~\eqref{eq:hier_l} using individual-event marginal likelihoods, $p(d_i|f_{0,i})$ for each event $i$, corresponding to either the homodyne or photon counting readouts.
{Our analysis uses events detected by CE above a standard threshold for the inspiral and merger portion of the waveform, thus the individual-event likelihood computation uses the known values of $M_i$ and $z_i$ to perform the mapping from event-level damped-sinusoid parameters into population parameters. Realistically, these parameter estimates will have some uncertainty, but the mass will be well-constrained at the 3\% relative uncertainty~\cite{HuxfordPRD24AccuracyNeutron}, equivalent to a frequency uncertainty of $3\%$ in Eq.~\eqref{eq:fpeak_eos}. The redshift will typically have a relative uncertainty of 10\%~\cite{SafarzadehA19MeasuringDelay}, equivalent to a frequency uncertainty of $<4\%$. Comparing these to the single-event credible intervals of Fig.~\ref{fig:snr_summary} indicates that they will negligibly widen the single-event posterior of $R_{1.6}$ for either readout scheme, so we adopt their exact values.}

We consider four noise configurations.
Until now, we have compared photon counting to an \textit{unsqueezed} homodyne - i.e. shown a direct comparison at constant SNR. 
While this is appropriate for a methodological comparison, the realistic alternative to photon counting is the homodyne design sensitivity which involves squeezing, 10\,dB is projected for CE, though its noise curve only achieves around 8\,dB above 1kHz\cite{MiaoPRX19QuantumLimit}.
The more fair comparison is therefore between photon counting with a design sensitivity classical noise (i.e. nominal operating power) and homodyne with design sensitivity. 
For comparison, we also explore the impact of a tenfold decrease in classical noise and and additional +5\,dB squeezing to show the effect of overcoming a statistical threshold in inference ability. {For the homodyne results at differing noise levels, we use the same noise realization but change its scaling. For the counting results at different noise levels, we use different noise realizations but consistent signal realizations.}

\begin{figure}
    \centering
    \includegraphics[width=\linewidth]{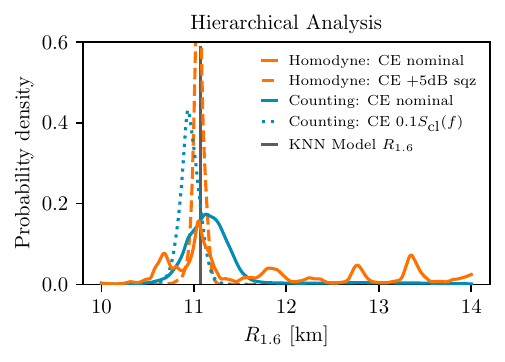}
    \caption{Marginal posterior distributions for the radius of a $1.6\,M_\odot$ NS with a homodyne (orange) and photon counting (blue) readout scheme. 
    The homodyne result {8}\,dB of squeezing above 1kHz (solid orange) is very broad and with secondary structure.
    The photon counting result with design sensitivity classical noise (solid blue) peaks at the correct value and yields a ${\sim}1\,$km radius measurement. 
    Achieving and additional 5\,dB of squeezing (dotted orange) or an order-of-magnitude reduction in the classical noise (dotted blue) tightens inference. The true value is shown with a vertical black line.}
    \label{fig:hierarchical}
\end{figure}

The posterior for $R_{1.6}$ is shown in Fig.~\ref{fig:hierarchical} with $10^4$ observations. 
Under the current design sensitivity, photon counting outperforms homodyne by about a factor of two.
The $68\%$ credible interval for $R_{1.6}$ is $1.25$\,km and $0.51$\,km for the homodyne with 10\,dB squeezing and photon counting with design sensitivity classical noise (and no squeezing), respectively. 
The homodyne posterior itself is very broad and with bimodal structure.
The significance of secondary maxima decreases with more observations or higher SNRs, so it is likely a statistical fluctuation due to the general lack of informative constraints. 
Furthermore, either increasing squeezing to 15\,dB for the homodyne readout or reducing the classical noise when photon counting leads to improved constraints of $0.07$\,km and $0.25$\,km respectively. The constraints with increased squeezing are driven by a few select observations. 
While this is an idealized demonstration of the method, we find that in the current design sensitivity of CE, photon counting boasts a ${\sim}2.5\times$ reduction in uncertainty in measuring $R_{1.6}$. 

\section{Implications}~\label{sec:imp}

We have explored photon counting as an alternative readout scheme of future GW detectors for bands dominated by quantum noise. 
Considering $>1\,$kHz data that are dominated by shot noise, we showed that photon counting can plausibly outperform traditional homodyne for {inferring the properties of BNS postmerger signals and constraining the EoS}. 
Sub-$1\,$kHz frequencies will be dominated by classical noise, hence the standard homodyne readout is preferable. 

The improved {inference} performance of the photon counting readout arises in part due to statistical inefficiencies in the analysis that arise from per-event marginalizations. {We have attempted to put the two readouts on equal statistical footing to compare them, while making realistic assumptions such as a lack of knowledge of precise phase modeling or knowledge of the merger time. We anticipate that further comparison or analysis is needed as modeling advances. Even if postmergers are slightly louder than expected and homodyne proves sufficient, one can straightforwardly apply conclusions of this analysis to related tasks (higher modes or measuring the EoS over time) where photon counting methods are more conclusively advantageous. Thus, we intend this work to help advance discussion of alternate readout methods and their potential statistical benefits.} {Moreover, current generation of GW detectors demonstrate that a large fraction of next-generation detector lifetime may operate below its projected sensitivity, where the advantage of photon counting over the standard readout is more clear. Furthermore, photon counting is impacted by the classical noise background, benefitting when that background is substantially lower than the quantum noise of the standard readout. This motivates continual improvement to GW detectors to reduce their classical noise backgrounds.}

{One takeaway from this analysis is that marginalized-vs-maximized parameters can affect hierarchical composition of near- and sub- threshold events. Increasing the SNR using detector tunings~\cite{SrivastavaA22SciencedrivenTunable, PhysRevD.99.102004} can thus have a nonlinear improvement to science output. Further studies into analyzing impacts of prior distribution to better understand the SNR threshold levels and rate of above/below threshold events may be useful future work to evaluate whether high-frequency tuned designs substantially improve posterior constraints in the nonlinear fashion indicated by Figs.~\ref{fig:snr_summary} and \ref{fig:hierarchical}.}

Photon counting has a number of advantages and disadvantages. 
First, the technology required to construct the temporal basis filters has not yet reached maturity and will rely on {considerable further} development of quantum memory and metrology apparatus to meet the {demanding needs of GW observatories}. 
We hope that this study further motivates development efforts.
Second, photon counting has a more limited domain of applicability than homodyne: quantum noise must dominate over classical noise and the expected signal must be simple enough that it can be reasonably expressed with a single- or few-photon count measurements from appropriately constructed filters. 
BNS postmergers signals are an ideal application as they are approximately composed of a collection of spectral peaks, appear in the detector $\gtrsim 1\,$kHz, and are mostly expected to remain below the detection threshold even for next-generation detectors. 
When all the conditions for photon counting are met, it can offer a quantifiable advantage. 
We show that photon counting could plausibly detect 1 BNS postmerger out of 100 signals with an SNR of 0.2. 
This then translates into improved EoS constraints over the whole BNS population.  

Beyond BNS postmerger signals, photon counting might prove efficient for high-frequency stochastic GW detection~\cite{Vermeulen:2024vgl, GuoPRD26FundamentalQuantum} where the homodyne detector response is low or for deviations from general relativity if appropriate filters can be motivated. 
Pursuing these options, along with a more systematic study of BNS postmerger inference {and detectability given a false-alarm rate}, will help clarify the capabilities of photon counting and motivate the development of this experimental readout scheme.

\section*{Acknowledgments}

We thank James Gardner, Sander Vermeulen, Ian MacMillan, Isaac Legred, Luis Lehner, Will Farr, Jacob Golomb, and Max Isi for thoughtful discussions on the research and helpful comments on the manuscript. 
EP was supported by NSF Grant PHY-2309200.
KC acknowledges support from the Department of Energy under award number DE-SC0023101.
LM acknowledges support from the 2024 Alfred P. Sloan foundation fellowship and NSF grants PHY-2317110 and PHY-2309268, respectively for the Institute of Quantum Information and Matter Physics Frontier Center and for the Cosmic Explorer optical design project.
The authors are grateful for computational resources provided by the LIGO Laboratory and supported by National Science Foundation Grants PHY-0757058 and PHY-0823459.

This analysis was built utilizing {\sc NumPyro}~\citep{phan2019composable, bingham2019pyro} and {\sc JAX}~\citep{jax2018github} for the inference calculation and data analysis. 
We leverage {\tt \sc AstroPy}~\cite{astropy:2013, astropy:2018, astropy:2022} and {\tt \sc SciPy}~\cite{Hunter:2007} for additional calculations, and {\tt \sc corner}~\cite{corner} and {\tt \sc arViz}~\cite{arviz_2019} for plotting purposes.
The code to replicate the results presented here is available at \url{https://github.com/ethanpayne42/GWPhotonCounting}.

\bibliography{refs}

\appendix

\section{Hierarchical inference without per-event time-marginalization}
\label{app:hierarchicalB}

\begin{figure}
    \centering
    \includegraphics[width=\linewidth]{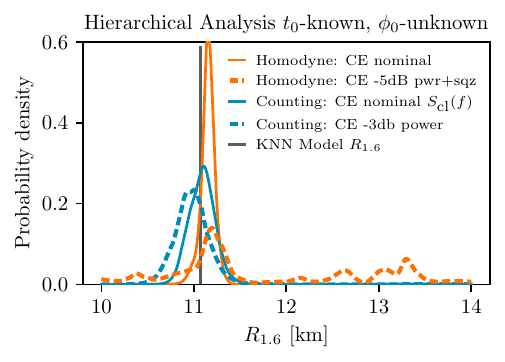}
    \caption{Marginal posterior distributions for the radius of a $1.6\,M_\odot$ NS with a homodyne and photon counting readout scheme. Here, the ringdown start time $t_0$ is assumed known and does not require marginalization.
      The homodyne result (solid orange) at nominal performance precisely estimates the radius, while a slightly degraded noise (-5db, coming from -3db in circulating power and -2db from squeezing) creates a broad posterior with secondary structure.
    The photon counting result with design sensitivity classical noise (solid blue) and with 3dB less arm power (dashed blue) peaks at the correct value and yields a ${\sim}0.5\,$km radius measurement. The true value is shown with a vertical black line.}
    \label{fig:hierarchicalB}
\end{figure}

{
For comparison with the hierarchical results of Sec.~\ref{sec:hier}, we repeat the calculation for $10^4$ events assuming that the postmerger start time can be accurately measured from the inspiral data and that future numerical relativity models at varying EoSs will be able to predict it. 
In practice this means that we keep the time fixed to its true value for each single-event analysis.
The aim of this exercise is to show the importance of the per-event marginalized parameters on the relative performance of the homodyne readout vs. photon counting. 
Figure~\ref{fig:hierarchicalB} shows the results, in which the nominal CE design can now constrain $R_{1.6}$ with its standard readout. This is in contrast to Fig.~\ref{fig:hierarchical} where the same detector configuration results in a wide $R_{1.6}$ posterior with multiple peaks when time is marginalized over.}

{This comparison demonstrates the influence of the marginalized parameters, by showing that now homodyne appears sufficient for the nominal CE noise projections. We include also an analysis where CE is 5db less sensitive. A plausible scenario for such reduced sensitivity can arise from two degradations: 3db less arm power (750kW), motivated by the current generation detectors (still) operating at less than half of their nominal power, and 2db less squeezing at high frequencies, which can arise if their signal extraction cavity has 1000ppm of internal optical loss rather than CE's current projection of 500ppm. The $-5$dB noise model of CE does not constrain $R_{1.6}$ well, while the reduced-arm-power photon counting (dashed blue) is still informative.}

{This comparison further supports photon counting as a potentially useful readout method in a plausible operating scenario for CE.}

\section{Fisher Information of Lorentzian Ringdowns}
\label{app:FI}

In this Appendix we calculate the Fisher Information for Lorentzian signals extracted with a homodyne readout scheme.
In GW analyses, the Fisher framework is most commonly invoked in the regime of large signal SNR where the likelihood is approximately Gaussian and the covariance is the Fisher inverse~\cite{Vallisneri:2007ev}.
Though most events are subthreshold and the per-event Fisher information does not represent each event's posterior, it still provides an upper limit on precision via the Cramer-Rao bound.
Additionally, we are interested in the asymptotic summation of the information over multiple events, which, by the Bernstein-von Mises theorem, can be related to the asymptotically-Gaussian width of hierarchical posterior in the limit of many events.

Specifically, we calculate the Fisher Information for common \emph{population-level} parameters such as the postmerger frequency (a proxy for the BNS EoS) for Lorentzian signals while incorporating \emph{event-level} nuisance parameters such as the phase. 
We focus on the phase as the nuisance parameter since complex merger physics, such rotation and  B-field dynamics, might make modeling it in terms of inspiral parameters impossible. 
The main paper analysis incorporates the assumption that the phase is unpredictable with a uniform prior both at the event and at the population level. 
Here, we show that this assumption changes how the Fisher Information about the frequency accumulates under homodyne, leaving room for a possible more favorable scaling with photon counting.

For simplicity of the computations, we assume \emph{complex} Lorentzian signals and a complex data stream, which reduces the algebra and allows for closed-form expressions. 
Below, $S_n$ is the two-sided power spectral density of the complex data stream, which is assumed to be white. 
Since the signal and noise are complex timeseries, the Fisher Information is generically twice the real case. One can remove this factor of two by alternatively interpreting the formulas below assuming that $S_n$ is a one-sided spectral density (2x larger than the two-sided) with real Lorentzian templates and at the regime of $\omega \gg \gamma$. Under that interpretation, these Fisher information formulas are representative of real Lorentzian signals on real noise despite being computed using complex signals and noise.

The complex Lorentzian ringdown signal is
\begin{equation}
  h(t| \boldsymbol{\theta}) = A e^{-\gamma t} e^{i(2\pi f t + \phi)} = A e^{i\phi} e^{-(\gamma - i\omega)t}\,,
\end{equation}
for $t>0$ and zero before. 
The parameter vector is $\boldsymbol{\theta} = (f, A, \phi, \gamma)$ with $f=\omega/2\pi$ the frequency, $A$ the amplitude, $\phi$ the phase, and $\gamma$ the damping rate.

\begin{widetext}
\subsection{Homodyne Likelihood Function}

The (log) likelihood for a model with parameters $\boldsymbol{\theta}$ and true signal parameters $\boldsymbol{\theta}_0$ assumes the standard Gaussian-noise form (ignoring constant terms):
\begin{align}
\log L(\boldsymbol{\theta}_0,\boldsymbol{\theta}) &= \frac{-1}{S_n}\int_0^{\infty} \Big|h(t | \boldsymbol{\theta}) - h(t | \boldsymbol{\theta}_0)  \Big|^2 \, \dd t\,,
  \\
  &= 2\text{Re}\left[\int_0^{\infty} \frac{AA_0 }{S_n} e^{-i(\phi-\phi_0)} e^{-(\gamma + \gamma_0 - i(\omega-\omega_0))t} \, dt\right] - \frac{A^2}{2\gamma S_n} - \frac{A_0^2}{2\gamma_0 S_n}\,, \label{eq:expand}
    \\
  &=  G \cos(\phi - \phi_0) + Q \sin(\phi - \phi_0) - \frac{A^2}{2\gamma S_n} - \frac{A_0^2}{2\gamma_0 S_n}\,,
\end{align}
where we define the coupling overlap functions $G$ and $Q$ that depend on everything but the phase 
\begin{align}
  G &=  \frac{2 AA_0}{S_n} \frac{(\gamma + \gamma_0)}{(2\pi\Delta f)^2 + (\gamma + \gamma_0)^2}, 
\qquad
  Q =  \frac{2 AA_0}{S_n} \frac{2\pi\Delta f}{(2\pi\Delta f)^2 + (\gamma + \gamma_0)^2}\,, 
\end{align}
and $\Delta f = f - f_0$. Note that the term in square brackets of Eq.~\eqref{eq:expand} is equivalent to Eq.~\eqref{eq:snr_olap}, thus $\textsc{SNR}^2 = G|_{\boldsymbol{\theta} = \boldsymbol{\theta}_0} = 2 A_0^2/S_n\gamma_0$. The following formulas can then be cast in terms of postmerger signal SNR metric.

Using the formulas above, the Fisher matrix is computed to be 
\begin{align}
  \mathcal{I}_{\boldsymbol{\theta}\boldsymbol{\theta}}
  &\equiv
  -
  \frac{\partial}{\partial \boldsymbol{\theta}^T}\frac{\partial} {\partial \boldsymbol{\theta}}
  \log L(\boldsymbol{\theta}, \boldsymbol{\theta}_0)\Big|_{\boldsymbol{\theta} = \boldsymbol{\theta}_0}
    = 
  \begin{bmatrix}
    \frac{2\pi^2 A_0^2}{\gamma_0^3 S_n} & 0 & -\frac{\pi A_0^2}{S_n \gamma_0^2} & 0 \\
    0 & \frac{1}{\gamma_0 S_n} & 0 &  \frac{A_0}{2 S_n \gamma_0^2} \\
    \frac{\pi A_0^2}{S_n \gamma_0^2} & 0 & \frac{A_0^2}{\gamma_0 S_n} & 0 \\
    0 & -\frac{A_0}{2 S_n \gamma_0^2} & 0 & \frac{A_0^2}{2\gamma_0^3 S_n}
  \end{bmatrix}.
\end{align}
Both terms involving the frequency (the $f{-}f$ and $f{-}\phi$ components) scale as $A_0^2/S_n$. 
The cross-terms between frequency and phase indicate that the two are correlated and uncertainty in the phase impacts the frequency.
The Fisher inverse still provides the Cramer-Rao Bound on the covariance matrix
\begin{align}\label{eq:FI_CRB}
  \mathcal{C}_{\boldsymbol{\theta}\boldsymbol{\theta}} \ge (\mathcal{I}^{-1})_{\boldsymbol{\theta}\boldsymbol{\theta}} \Rightarrow \mathcal{C}_{ff}
  \ge
  \left(\frac{\pi^2 A_0^2}{\gamma_0^3 S_n}\right)^{-1}
  \ge 
 (\mathcal{I}_{ff})^{-1}\,.
\end{align}
which scales linearly with the noise PSD.

The relevance of the Fisher Information matrix of each event towards a population analysis is to sum all of the information after changing variables from the per-event $f$ to the population-parameter $R_{1.6}$ (this change of variables is not done here for our purpose of illustration). 
The total/aggregate information then constrains the width of the posterior through the Cramer-Rao Bound, i.e., if the computed aggregate information is low compared to some useful posterior width-scale for relevant population parameters, then a hierarchical Bayesian analysis simply cannot and will not produce a useful posterior.

The above per-event Fisher, however, is not indicative of how the population inference posterior scales.
The key issue is that the \emph{entire vector} of parameters are set to their nominal value $\boldsymbol{\theta} = \boldsymbol{\theta}_0$ when evaluating the information. 
The underlying assumption is that in any asymptotic estimate over a population of events (i.e., a hierarchical analysis), the phase of the signal becomes (well-) known. 
However, since each and every event is expected to have an \emph{unknown} phase, the assumption of asymptotic knowledge of the phase is incorrect. One can try to work around this logic by using a Schur-compliment formula to produce an FI matrix without the phase parameter. This procedure will produce the same bound as Eq.~\eqref{eq:FI_CRB} by turning the last inequality into an equality (the inverse on the original Fisher matrix is internally applying the same formula).

However, the Bernstein-von Mises theorem has several technical conditions. 
The relevant condition for this discussion is that the total number of parameters of the posterior distribution must be asymptotically finite. 
When the phase is estimated individually on each event, the number of parameters scales with the number of events and this condition of the theorem is violated, when using a Schur-reduced FI matrix with the nuisance parameters removed. 

The Fisher information becomes useful and relevant if the phase marginalization is included in the likelihood used to compute the Fisher Information, in which case it can no longer be included one of the parameters of the matrix and cannot be interpreted as one of the parameters of the posterior.
Using a marginalized likelihood for the Bayesian update is equivalent to including the parameter at each event and marginalizing the posterior.
While the two cases are equivalent in a Bayesian hierarchical analysis, they are not equivalent for the formulation and interpretation of the Fisher information through the Bernstein-von Mises theorem due to this technical requirement.

Thus, to fix the formulation, we adjust the likelihood function to incorporate the per-event averaging over the nuisance parameter's prior distribution (uniform, for the case of the phase parameter).

\subsection{Homodyne Likelihood Function with phase marginalization}

In this formulation we adopt a reduced parameter set $\overline{\boldsymbol{\theta}} = (f, A, \gamma)$ and marginalize over the phase:
\begin{align}
   \log  L(\overline{\boldsymbol{\theta_0}}, \overline{\boldsymbol{\theta}})
 &= \log\left( \frac{1}{2\pi}\int_0^{2\pi} \exp\left(\frac{-1}{S_n}\int_0^{\infty} \Big|h(t | \boldsymbol{\theta}) - h(t | \boldsymbol{\theta}_0)  \Big|^2 \, dt  \right) d\phi\right)\\
 &=\log\left(\frac{1}{2\pi} \int_0^{2\pi} e^{G \cos(\phi-\phi_0) + Q\sin(\phi-\phi_0) } d\phi \right) - \frac{A^2}{2\gamma S_n} + \frac{A_0^2}{2\gamma_0 S_n}\,.
\end{align}
Frequency estimation depends only on the first term
\begin{align}
\log \overline{L} &\equiv \log\left(\frac{1}{2\pi} \int_0^{2\pi} e^{G \cos(\phi) + Q\sin(\phi) } d\phi \right)\,.
\end{align}

\subsubsection{Large Amplitude Signals with phase-marginalization}

When $A_0^2/S_n \gg 1$, signals are in a high-SNR regime where all parameters are determined and $\phi\approx \phi_0$. 
The likelihood can be evaluated by expanding the phase (in what follows we set $\phi_0$ for simplicity) and using Gaussian integrals, giving
\begin{align}
  \log \overline{L}
  &\approx
    \log\left(\frac{1}{2\pi} \int_{-\infty}^{\infty} e^{G + Q\phi- G\frac{\phi^2}{2}} d\phi \right)
    =
    \log\left(\frac{1}{2\pi} \int_{-\infty}^{\infty} e^{G + \frac{Q^2}{2G} - \frac{G}{2}(\phi - \frac{Q}{G})^2} d\phi \right)
  \\ &
       \approx
G + \frac{Q^2}{2G} - \log(2\pi) + \log\left( \int_{-\infty}^{\infty} e^{-\frac{G\phi^2}{2}} d\phi \right)
  \\ &
       \approx
G + \frac{Q^2}{2G} - \log(2\pi) + \log\left( \sqrt{\frac{2\pi}{G}}\right)
       \approx
G + \frac{Q^2}{2G} - \frac{\log(G)}{2} - \frac{\log(2\pi)}{2}\,.
\end{align}
In this case, the Fisher Information (and the corresponding Cramer-Rao bound) can be shown to be identical to the 4-parameter case without phase marginalization:
\begin{align}
  \mathcal{C}_{ff} \geq \mathcal{I}^{-1}_{ff} 
  = \left(\frac{\pi^2 (A_0^2 - S_n\gamma_0)}{\gamma_0^3 S_n}\right)^{-1} \approx \left(\frac{\pi^2 A_0^2 }{\gamma_0^3 S_n}\right)^{-1} = \frac{2}{\pi^2}\frac{\gamma_0^2}{\textsc{Snr}^2}.
\end{align}
Because in the high-SNR case the phase is well determined, the Fisher Information replicates the scaling of the full 4-parameter inference, where phase does not need to be estimated because it is well known asymptotically.

\subsubsection{Small Amplitude Signals with phase-marginalization}

When $G$ and $Q$ are small, we instead expand the exponential to second order:
\begin{align}
  \log \overline{L}
  &\approx
    \log\left( \frac{1}{2\pi}\int_{-\pi}^{\pi} 1 + G \cos\phi + Q \sin\phi + \frac{G^2}{2} \cos^2\phi + \frac{Q^2}{2} \sin^2\phi + GQ \sin\phi \cos\phi\, d\phi \right)
  \\ &
       \approx \log(1 + G^2/4 + Q^2/4)\,.
\end{align}
and calculate the Fisher Information for frequency estimation in the low SNR, $A_0^2/S_n \ll 1$, regime
\begin{align}
  \mathcal{C}_{ff} \geq\mathcal{I}^{-1}_{ff} = \left[\frac{\pi^2 A_0^4}{S_n^2 \gamma_0^4\left(1 + \frac{A_0^4}{4S_n^2 \gamma_0^2}  \right)}\right]^{-1} \approx \left(\frac{\pi^2 A_0^4}{S_n^2 \gamma_0^4}\right)^{-1}
 = \frac{4}{\pi^2}\frac{\gamma_0^2}{\textsc{Snr}^4}\,.
\end{align}
The Fisher Information now scales as $A_0^4/S_n^2$, compared to $A_0^2/S_n$ in the high-SNR/known-phase case.
The phase being unknown causes a much worse scaling of the information.
Since the Fisher Information is still an upper bound on the precision with which frequency can be estimated, the true scaling of hierarchical inference from many subthreshold events will be \emph{at best} $A_0^4/S_n^2$, in practice probably even less favorable.

The change in the Fisher Information scaling is not unexpected because an unknown phase prevents any hierarchical analysis from coherently adding the signals to average away the noise. 
Instead, the full analysis scales similarly to excess-power type analyses, whose estimators typically have the scaling of $A_0^4/S_n^2$ in their detection or estimation convergence rate.
This is true for the frequency even though it is itself shared among all events (or at least it is deterministic at the population level).
The frequency's correlation with random per-event parameters deteriorates its information scaling to that of inherently random parameters.

The analysis of the main text takes advantage of this unfavorable scaling to switch to photon counting which is shown to outperform homodyne in this regime as it is more optimal for  incoherent/stochastic signal estimation. 
The major result of this work is that this alternate quantum measurement improves the estimation of a nonrandom population-level parameter due to the randomness of other event-level parameters that must be marginalized. 
\end{widetext}

\end{document}